\documentclass{ieeeaccess}
\usepackage[none]{hyphenat}
\usepackage[acronym,nonumberlist]{glossaries}
\newacronym{AC}{AC}{Access Categorie}
\newacronym{ACK}{ACK}{Acknowledgement}
\newacronym{ACU}{ACU}{Admission Control Unit}
\newacronym{AF-HCCA}{AF-HCCA}{Adaptive Feedback-based HCCA Scheduler}
\newacronym{AID}{AID}{Associate Identifier}
\newacronym{AIFS}{AIFS}{Arbitration Inter Frame Space}
\newacronym{AMC}{AMC}{Adaptive Modulation and Coding}
\newacronym{AMTXOP}{AMTXOP}{Adaptive Multipolling TXOP Scheme}
\newacronym{AP}{AP}{Access Point}
\newacronym{ATXOP}{ATXOP}{Adaptive TXOP Scheme}
\newacronym{AWK}{AWK}{Alfred Aho, Peter Weinberger}
\newacronym{BSS}{BSS}{Basic Service Set}
\newacronym{CA}{CA}{Collision Avoidance}
\newacronym{CAP}{CAP}{Controlled Access Phase}
\newacronym{CBR}{CBR}{Constant Bit Rate}
\newacronym{CDF}{CDF}{Complementary Cumulative Distribution Function}
\newacronym{CF}{CF}{Contention Free}
\newacronym{CFP}{CFP}{Contention Free Period}
\newacronym{CF-Poll}{CF-Poll}{Contention Free}
\newacronym{CP}{CP}{Contention Period}
\newacronym{CSMA}{CSMA}{Carrier Sense Multiple Access}
\newacronym{CW}{CW}{Contention Window}
\newacronym{CWmax}{CWmax}{Contention Window Maximum}
\newacronym{CWmin}{CWmin}{Contention Window Minimum}
\newacronym{DCF}{DCF}{Distributed Coordination Function}
\newacronym{DEB}{DEB}{Deterministic Back-off}
\newacronym{DSSS}{DSSS}{Direct Sequence Spread Spectrum}
\newacronym{DTB}{DTB}{Dual Token Bucket}
\newacronym{DTH}{DTH}{Dynamic TXOP HCCA}
\newacronym{EDCA}{EDCA}{Enhanced Distributed Channel Access}
\newacronym{EDCF}{EDCF}{Enhanced DCF}
\newacronym{EDD}{EDD}{Earliest Due Date}
\newacronym{FFBI}{FFBI}{Feed-Forward Bandwidth Indication}
\newacronym{F-Poll}{F-Poll}{Feasible Polling Scheme}
\newacronym{FTP}{FTP}{File Transfer Protocol}
\newacronym{GI}{GI}{Generation Interval}
\newacronym{HC}{HC}{Hybrid Coordinator}
\newacronym{HCCA}{HCCA}{HCF Controlled Channel Access}
\newacronym{HCF}{HCF}{Hybrid Coordination Function}
\newacronym{HRDSSS}{HR/DSSS}{High-Rate DSSS}
\newacronym{HTTP}{HTTP}{Hypertext Transfer Protocol}
\newacronym{IEEE}{IEEE}{Institute of Electrical and Electronics Engineers}
\newacronym{IP}{IP}{Internet Protocol}
\newacronym{IR}{IR}{Infrared}
\newacronym{ISM}{ISM}{Industrial, Science, Medical}
\newacronym{LRE}{LRE}{Limited Relative Error}
\newacronym{MAC}{MAC}{Medium Access Control}
\newacronym{MSDU}{MSDU}{MAC-Service Data Unit}
\newacronym{MSI}{MSI}{Maximum Service Interval}
\newacronym{NAV}{NAV}{Network Allocation Vector}
\newacronym{NPHCCA}{NPHCCA}{Non-Polling based HCCA}
\newacronym{OFDM}{OFDM}{Orthogonal Frequency Division Multiplexing}
\newacronym{OSI}{OSI}{Open Systems Interconnection}
\newacronym{PC}{PC}{Point Coordinator}
\newacronym{PCF}{PCF}{Point Coordination Function}
\newacronym{PER}{PER}{Packet Error Rate}
\newacronym{PF}{PF}{Persistence Factor}
\newacronym{PHY}{PHY}{Physical Layer mode}
\newacronym{PIFS}{PIFS}{PCF Inter Frame Space}
\newacronym{PLCP}{PLCP}{Physical Layer Convergence Procedure}
\newacronym{QAP}{QAP}{QoS-enabled Acces Point}
\newacronym{QBSS}{QBSS}{QoS-supporting Basic Service Set}
\newacronym{QoS}{QoS}{Quality of Service}
\newacronym{QS}{QS}{Queue Size}
\newacronym{QSTA}{QSTA}{QoS-enabled Station}
\newacronym{RAM}{RAM}{Random Access Memory}
\newacronym{RF}{RF}{Radio Frequency}
\newacronym{RM}{RM}{Rate Monotonic}
\newacronym{RTCP}{RTCP}{RTP Control Protocol}
\newacronym{RTP}{RTP}{Real-time Transport Protocol}
\newacronym{RTS/CTS}{RTS/CTS}{Request to Send/Clear to Send}
\newacronym{SI}{SI}{Service Interval}
\newacronym{SIFS}{SIFS}{Short Inter Frame Space}
\newacronym{SINR}{SINR}{Signal-to-Interference Plus Noise Ratio}
\newacronym{SMA}{SMA}{Simple Moving Average}
\newacronym{SMTP}{SMTP}{Simple Mail Transfer Protocol}
\newacronym{STA}{STA}{Station}
\newacronym{TBTT}{TBTT}{Target Beacon Transmission Time}
\newacronym{TC}{TC}{Traffic Category}
\newacronym{TCP}{TCP}{Transmission Control Protocol}
\newacronym{TGe}{TGe}{IEEE802.11 Task Group E}
\newacronym{TS}{TS}{Traffic Stream}
\newacronym{TSPEC}{TSPEC}{TS Specification}
\newacronym{TXOP}{TXOP}{Transmission Opportunity}
\newacronym{UDP}{UDP}{User Datagram Protocol}
\newacronym{UGC}{UGC}{User-Generated Content}
\newacronym{UP}{UP}{User Priorities}
\newacronym{VBR}{VBR}{Variable Bit Rate}
\newacronym{VoWLAN}{VoWLAN}{Voice over WLAN}
\newacronym{WLAN}{WLAN}{Wireless Local Area Network}
\newacronym{WM}{WM}{Wireless Medium}

\usepackage{graphicx}
\usepackage{amsmath, amsthm, amssymb}
\usepackage{multirow}
\usepackage{epstopdf}
\usepackage{multirow}   
\usepackage{booktabs}   
\usepackage[figuresright]{rotating}
\usepackage{array}
\usepackage{datatool}
\usepackage{longtable} 
\usepackage{soul}
\usepackage[lined, linesnumbered, ruled]{algorithm2e}
\usepackage{xcolor}

\usepackage{textcomp}
\def\BibTeX{{\rm B\kern-.05em{\sc i\kern-.025em b}\kern-.08em T\kern-.1667em\lower.7ex\hbox{E}\kern-.125emX}}

\begin{document}
\history{Date of publication xxxx 00, 0000, date of current version xxxx 00, 0000.}
\doi{00.0000/ACCESS.2018.DOI}

\title{Review on QoS provisioning approaches for supporting video traffic in IEEE802.11e: challenges and issues}
\author{\uppercase{Mohammed A. Al-Maqri}\authorrefmark{1,2},
\uppercase{Mohamed A. Alrshah}\authorrefmark{1},\IEEEmembership{Senior Member, IEEE},
\uppercase{and Mohamed Othman}.\authorrefmark{1,3},\IEEEmembership{Senior Member, IEEE}}

\address[1]{Department of Communications Technology and Networks, Universiti Putra Malaysia, Serdang, 43400, Malaysia}
\address[2]{Azal University for Human Development, 60th Street, Sana'a, Yemen}
\address[3]{Computational Science and Mathematical Physics Lab, INSPEM, Universiti Putra Malaysia, Serdang, 43400, Malaysia}

\tfootnote{This research was funded by the Malaysian Ministry of Education under UPM/700-2/1/GPB/2017/9557900 Putra with High-Impact Grant.}

\markboth
{M. A. Al-Maqri \headeretal: Review on QoS provisioning approaches for supporting video traffic in IEEE802.11e}
{M. A. Al-Maqri \headeretal: Review on QoS provisioning approaches for supporting video traffic in IEEE802.11e}

\corresp{Corresponding author: Mohamed A. Alrshah (e-mail: mohamed.asnd@gmail.com)}

\begin{abstract}
Recently, the demand for multimedia applications is dramatically increased, which in turn increases the portion of video traffic on the Internet. The video streams, which require stringent Quality of Service (QoS), are expected to occupy more than two-thirds of web traffic by 2019. IEEE802.11e has introduced HCF Controlled Channel Access (HCCA) to provide QoS for delay-sensitive applications including highly compressed video streams. However, IEEE802.11e performance is hindered by the dynamic nature of Variable Bit Rate (VBR) video streams in which packet size and interval time are rapidly fluctuating during the traffic lifetime. In order to make IEEE802.11e able to accommodate with the irregularity of VBR video traffic, many approaches have been used in the literature. In this article, we highlight and discuss the QoS challenges in IEEE802.11e. Then, we classify the existing QoS approaches in IEEE802.11e and we also discuss the selection of recent promising and interesting enhancements of HCCA. Eventually, a set of open research issues and potential future directions is presented.
\end{abstract}


\titlepgskip=-15pt

\maketitle

\section{Introduction}\label{sec:intro}
\IEEEPARstart{T}{he} optimal transport of delay-constrained multimedia services over WLANs requires adaptation to many aspects of \gls{OSI} model layers starting from delay constraints and bandwidth variations of the traffic at the application layer up to accommodation to wireless channel conditions and power constraints at the physical layer. The efficiency of 802.11e \gls{HCCA} function mainly depends on the accuracy of its scheduler in assigning network resources, such as channel bandwidth, to the traffic streams without jeopardizing the QoS constraints such as delay and throughput. Moreover, with the presence of delay-sensitive multimedia traffic with variable profile, the existing scheduling approaches become inefficient. Thus, the scheduler is required to consider the fluctuation of traffic in the scheduling process. 

This article introduces an overview of the prime challenges for provisioning QoS for multimedia traffic with emphasize on \gls{VBR} traffic in IEEE~802.11e wireless networks. Then, it presents a taxonomy for the existing solutions, and describes the most representative properties, advantages, and design challenges. This taxonomy comprises the core approaches and techniques on IEEE802.11e protocol, with more emphasize on \gls{HCCA} enhancements. Additionally, a systematic summarization and comparison for research contributions in each field are used to clearly identify the current challenges for further research. Finally, the article discusses the most critical issues which hinder the provisioning of QoS in wireless networks with a special attention to polling and \gls{TXOP} allocation enhancements.

This paper is a survey of QoS provisioning for video transmission in IEEE802.11e, which is organized as follows: Section~\ref{sec:11eStandard} exhibits the background about IEEE802.11e standard and its functions. Section~\ref{sec:QoSChalng} presents the main challenges in IEEE802.11e WLANs. Section~\ref{sec:classification} classifies and reviews the core approaches in IEEE802.11e WLANs, which were proposed to enhance QoS provisioning for multimedia traffic. A number of leading approaches aiming at improving the QoS for multimedia traffic has been discussed in Section \ref{sec:HCCA_enhancement}. Section~\ref{sec:comparison} shows a general comparison of the IEEE802.11e approaches and their targeted features, and lists some of the strength and limitation criteria of these approaches. Section~\ref{sec:ORI} and Section~\ref{sec:futureWork} identifies research trends, challenges, and potential future areas related to the article's scope, and finally Section~\ref{sec:conclusion} concludes the article.

\section{IEEE802.11e standard}
\label{sec:11eStandard}
Several amendments have been made to the legacy IEEE802.11 WLAN standard \cite{IEEEStand1999}, as shown in Table~\ref{tab:IEEEFamily}. IEEE802.11e is one of the approved versions of IEEE802.11 standard, which defines a combination of \gls{QoS} improvements on the \gls{MAC} layer for WLAN applications, as shown in Fig. \ref{fig001}. The standard is critically important for applications that are very sensitive to delay, such as \gls{VoWLAN} and multimedia streaming.
\begin{table*}[!t]
	\caption {The family of IEEE-802.11 versions}
	\centering
	\begin{tabular}{p{2.2cm}p{8cm}p{6cm}}
		\hline
		Standard 								& Objective 																& Frequency and Modulation\\ \hline
		IEEE802.11  \cite{IEEE80211Stand} 		& To provide up to 2 Mbps bit rate. 										& 2.4 GHz by utilizing DSSS and FHSS. \\
		IEEE802.11a \cite{IEEE80211aStandard}	& To provide up to 54 Mbps bit rate.										& 5 GHz by utilizing OFDM. \\
		IEEE802.11b \cite{IEEE80211bStandard} 	& To provide up to 11 Mbps bit rate.										& 2.4 GHz by utilizing HRDSSS. \\
		IEEE802.11c \cite{IEEESTD11c}			& To ensures proper bridging operations.									& -\\
		IEEE802.11d \cite{IEEE80211d}			& To covers more regulatory domains.										& -\\
		IEEE802.11e \cite{IEEE80211e2005} 		& To define new QoS enhancements to 802.11a and 802.11b. 					& -\\
		IEEE802.11f \cite{IEEE80211f}			& To provide interoperability for roaming among different APs.				& -\\
		IEEE802.11g \cite{IEEE80211g} 			& To provide up to 54 Mbps bit rate.										& 2.4 GHz by utilizing OFDM. \\
		IEEE802.11n \cite{IEEE80211n} 			& To provide up to 600 Mbps bit rate.										& 2.4 and 5 GHz by utilizing MIMO-OFDM.\\ \hline
	\end{tabular}
	\label{tab:IEEEFamily}
\end{table*}

In IEEE802.11e,  the QoS feature includes an extra coordination function called \gls{HCF}. This function combines both functionalities of the well-known \gls{PCF} and \gls{DCF}. In order to permit the use of a uniform assortment of frame exchange sequences for QoS data transfers during the time of both \gls{CP} and \gls{CFP}, the HCF introduced some enhanced frame subtypes and QoS-specific mechanisms. As for contention-based transfer, HCF employs a contention-based channel access approach, namely \gls{EDCA}, while for contention-free transfer it uses a controlled-channel access method, so-called \gls{HCCA}. \glspl{STA} might obtain TXOPs using \gls{EDCA}, \gls{HCCA} or both schemes together. Thus, a \gls{TXOP} is defined as \gls{EDCA} \gls{TXOP} if it is obtained by the contention-based channel access, while it is defined as \gls{HCCA}-\gls{TXOP} if it is obtained by the controlled channel access.

\Figure[h!](topskip=0pt, botskip=0pt, midskip=0pt)[width=0.9\linewidth]{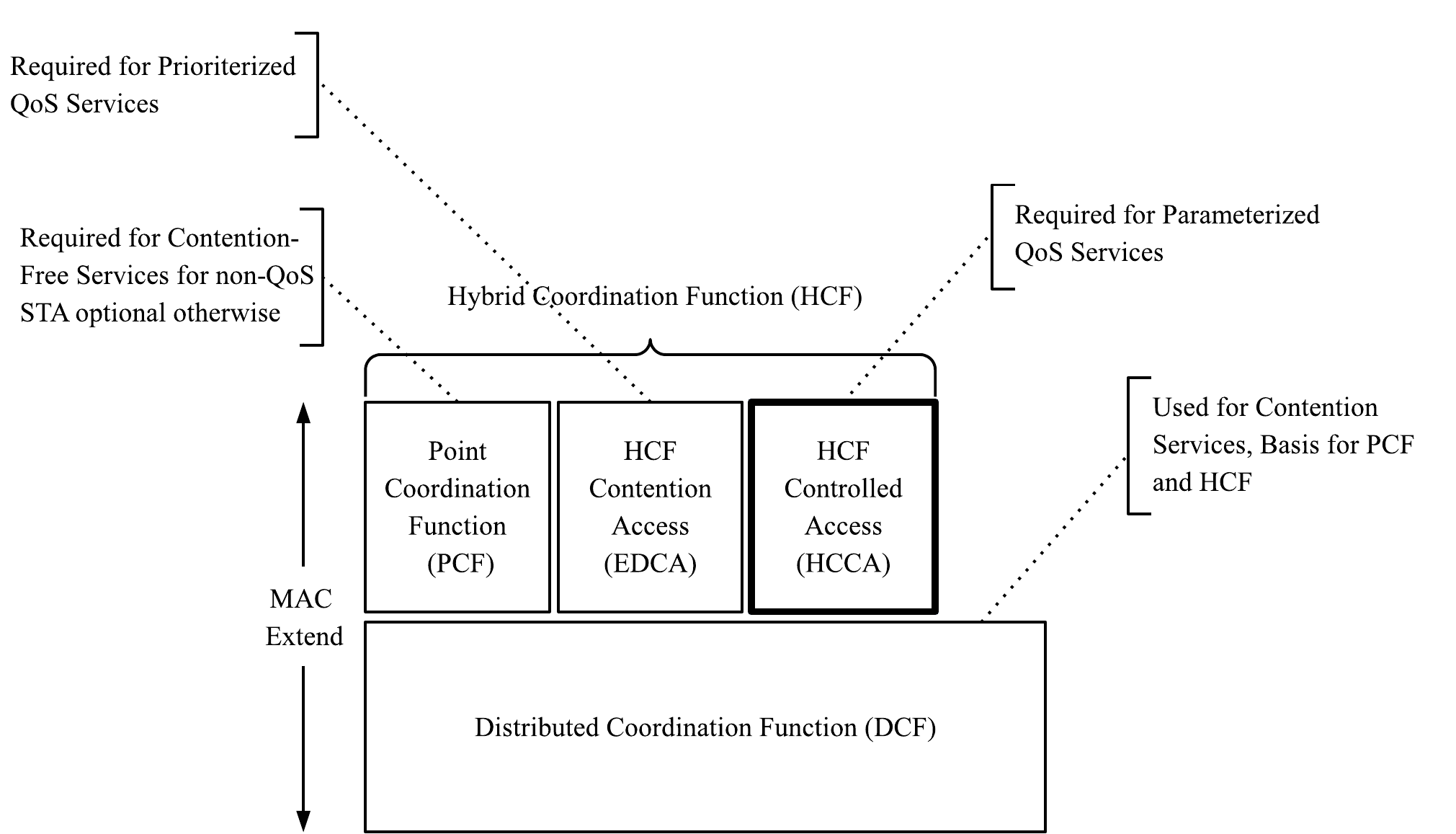}
{MAC architecture in IEEE802.11e.\label{fig001}}

\subsection{Enhanced distributed channel access (EDCA)}
\gls{EDCA} mechanism has been designed to provide sort of differentiated distributed access to \gls{WM} for \glspl{STA} by using eight uneven \glspl{UP}. It determines four \glspl{AC} to provide support for traffic delivery at the \glspl{STA} using \glspl{UP}, which produces the \gls{AC}, as shown in Table~\ref{tab:ACMap}. For every \gls{AC}, an enhanced variant of \gls{DCF}, called \gls{EDCF}, contends for TXOPs using a set of \gls{EDCA} parameters. For more details about the \gls{EDCF} refer to \cite{IEEEStandard2012}. Implementation of this mechanism is easy; however, the QoS requirement of a realtime traffic can not always be met, especially when the heavy load conditions occur. In heavy loaded scenarios, higher prioritized traffic QoS requirement may easily be broken even though it exhausts most of the available bandwidth. However, lower prioritized traffic may be starved and severely deteriorated in both efficiency and effectiveness.
\begin{table} 
	\centering
	\caption {Mappings of user priority to access category}	
	\begin{tabular}{p{2cm}p{1.5cm}p{1.5cm}p{2cm}}
		\hline
		Priority	& UP	& AC 		& Designation  	\\ \hline
		Lowest		& 1 	& AC\_BK 	& Background 	\\
					& 2 	& AC\_BK 	& Background 	\\
					& 0 	& AC\_BE 	& Best Effort	\\
					& 3 	& AC\_BE 	& Best Effort	\\
					& 4 	& AC\_VI 	& Video 		\\
					& 5 	& AC\_VI 	& Video 		\\
					& 6 	& AC\_VO 	& Voice 		\\
		Highest		& 7 	& AC\_VO 	& Voice 		\\ \hline
	\end{tabular}
	\label{tab:ACMap}
\end{table}

\subsection{HCF controlled channel access (HCCA)}
\begin{figure}[h!]
	\centering
	\includegraphics[width=0.9\linewidth]{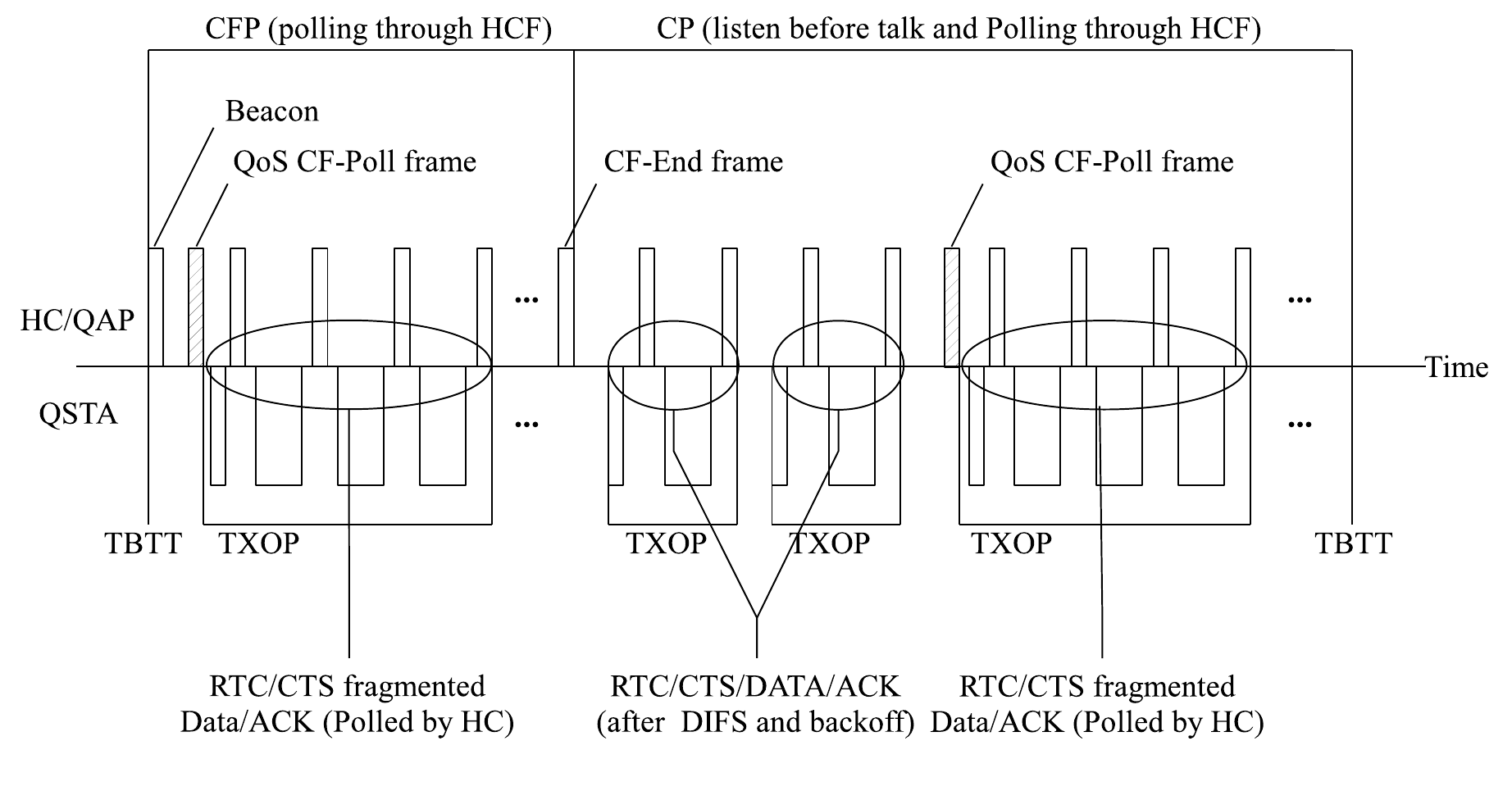}
	\caption{An 802.11e super-frame example, CFP and CP. In the CFP, the frame exchange takes a place throughout the polling mechanism, while in CP the QSTAs have to listen to the medium transmitting data packets.}
	\label{fig:MACArch}
\end{figure}

As known in IEEE-802.11e, a synchronization signal is rhythmically sent to all of the connected stations in the \gls{BSS}. The time between two subsequent signals makes a super-frame, where a service can be delivered through this super-frame over two periods of time, \gls{CFP} and \gls{CP}. The data of any station has to be transmitted during a period of time, namely \gls{TXOP}, which is dedicated for a \gls{QSTA} to transfer its \glspl{MSDU}. Fundamentally, \gls{TXOP} is acquired through the contention-based access, which is known as EDCA-TXOP. As for the controlled medium access, the \gls{HC} grants the \gls{TXOP} to the \gls{QSTA} (known as polled \gls{TXOP}). Fig. \ref{fig:MACArch} shows a clear example of 802.11e super-frame which demonstrates the interchanging of one controlled medium access and one contention-based period, where the later includes one \gls{QAP} and three \glspl{QSTA}. In general, controlling medium access occurs either within the \gls{CP} or through the \gls{CFP} if the medium remains idle for at least one period of \gls{PIFS}. In order to support \gls{QoS} in \gls{HCCA}, many researchers have proposed to improve the existing \gls{PCF} by controlling the transmission only within the \gls{CFP}. Therefore, the data packets of any wireless station in \gls{HCCA} can be only transmitted during a declared period of time in the poll frame.

\subsubsection{Reference design of \gls{HCCA}}
At the point if a \gls{QSTA} wants to transmit its realtime \gls{TS} within the contention-free period, it has to send an ADDTS-Request to the \gls{QAP}. This ADDTS-Request declares the requirements of QoS for that specific \gls{TS} within the relevant \gls{TSPEC} domain. Consequently, the \gls{QAP} will try to fulfill the requirements while conserving the QoS of existing admitted flows. If the ADDTS-Request is accepted, the \gls{QAP} will reply an ADDTS-Response back to the relevant station, then, this station will be admitted to the \gls{QAP} polling list. Table~\ref{tab:TSPECsymbol} shows the compulsory \gls{TSPEC} parameters and their symbols.

\begin{table} [!h]
	\caption {Symbols used for \gls{TSPEC} and scheduling parameters}
	\centering
	\begin{tabular}{p{2cm}p{6cm}}
		\hline
		Notation& Description\\ \hline
		$\rho$	& Mean Data Rate\\
		$L$ 	& Nominal \gls{MSDU} Size \\
		$M$		& Maximum \gls{MSDU} Size \\
		$D$ 	& Delay Bound\\
		$SI$ 	& Service Interval\\
		$mSI$ 	& Minimum Service Interval\\
		$MSI$ 	& Maximum Service Interval\\
		$R$ 	& Physical Transmission Rate\\
		$BI$	& Beacon Interval \\
		$O$		& \gls{PHY} and \gls{MAC} Overhead \\
		$N$		& Number of packets \\
		$T$		& super-frame duration \\
		$T_{CP}$ & Contention-based duration \\ \hline
	\end{tabular}
	\label{tab:TSPECsymbol}
\end{table}

After accepting new ADDTS-Request, the \gls{HCCA} scheduler will go through the following steps:
\begin{enumerate}
	\item \textit{Assigning service interval} \\
	\label{SIassign}
	\gls{HCCA} \cite{IEEEStandard2012} computes the \gls{SI} as a sub-multiple of the whole Beacon Interval $BI$, which is calculated as the minimum of the maximum \glspl{SI} of all priorly accepted traffic streams including the incoming data traffic. Equation (\ref{si}) is used to calculate the \gls{SI}:
	\begin{equation}
	\label{si}
	SI = \frac{BI}{\left \lceil\frac{BI}{MSI_{min}} \right \rceil},
	\end{equation}
	where $MSI_{min}$ is computed as in Equation (\ref{minSI}):
	\begin{equation}
	MSI_{min} = min(MSI_{i}) , i \in [1, n],
	\label{minSI}
	\end{equation}
	where $MSI_{i}$ denotes the maximum $SI$ of the $i^{th}$ stream and $n$ denotes the number of all previously admitted \glspl{QSTA}' traffic streams.
	
	\item \textit{Allocating TXOP} \\
	Variant \gls{TXOP} is allocated by \gls{HC} to every accepted \gls{QSTA} based on the declared QoS parameters in the \gls{TSPEC}, which allows the \gls{QSTA} to obtain the required QoS. The HC calculates \gls{TXOP} for the $i^{th}$ \gls{QSTA} based on the expected MSDUs, which may arrive at $\rho_{i}$, as calculated in Equation~\eqref{eq:N}:
	\begin{equation}
	N_{i}=\left \lceil \frac{SI\times\rho_{i}}{L_{i}} \right \rceil,
	\label{eq:N}
	\end{equation}
	where $L_{i}$ denotes the \gls{MSDU} of the $i^{th}$ station.\\
	
	Thereafter, the \gls{TXOP} of the $i^{th}$ station $(TXOP_{i})$ is calculated as the required time to transmit $N_{i}$ \gls{MSDU} or one maximum \gls{MSDU} at the relevant physical rate $R_{i}$, as in Equation~\eqref{eq:TXOP} below:
	\begin{equation}
	TXOP_{i}=max\left (\frac{N_{i} \times L_{i}}{R_{i}} + O, \frac{M}{R_{i}} + O \right)
	\label{eq:TXOP}
	\end{equation}
	where $O$ represents the total overhead, including MAC and physical headers, poll frames overheads, inter-frame spaces (IFSs) and acknowledgments.
	
	\item \textit{Admission control} \\
	The \gls{ACU} regulates the admission of the \gls{TS} while maintaining the QoS of the previously admitted \glspl{TS}. 
	When the \gls{ACU} receives a request of admitting a new \gls{TS}, the \gls{ACU} calculates a new $SI$ using Equation \eqref{si} and estimates the number of MSDUs that may arrive at this new $SI$ based on Equation \eqref{eq:N}. Then, the \gls{ACU} calculates the $TXOP_{i}$ for the particular \gls{TS} using Equation~\eqref{eq:TXOP}. Finally, the \gls{ACU} would admit the relevant \gls{TS} only if the following inequality is satisfied:
	\begin{equation}
	\label{eq:ACU}
	\frac{TXOP_{n+1}}{SI}+\sum_{i=1}^{n} \frac{TXOP_{i}}{SI}\leq \frac{T- T_{CP}}{T}
	\end{equation}		
	Fig.~\ref{fig:ACU} shows an example of an admitted stream from $STA_{i}$. The beacon interval is 100ms and the maximum \gls{SI} for the stream is 60ms. The scheduler sets a scheduled $SI$ to 50ms with complying to Equation~\eqref{eq:ACU}, where $n$ represents the number of all admitted streams, $n+1$ denotes the index of incoming \gls{TS}, $T$ indicates the beacon interval and $T_{CP}$ is the time reserved for \gls{EDCA} contention-period. 
	
	The \gls{HC} sends an ADDTS-Response to the relevant \gls{QSTA} only if Equation \eqref{eq:ACU} is satisfied, and it sends a message of rejection otherwise. Then, the \gls{HC} will add the accepted \gls{TS} to its polling list.	
	\begin{figure}[!h]
		\centering
		\includegraphics[width=0.8\linewidth]{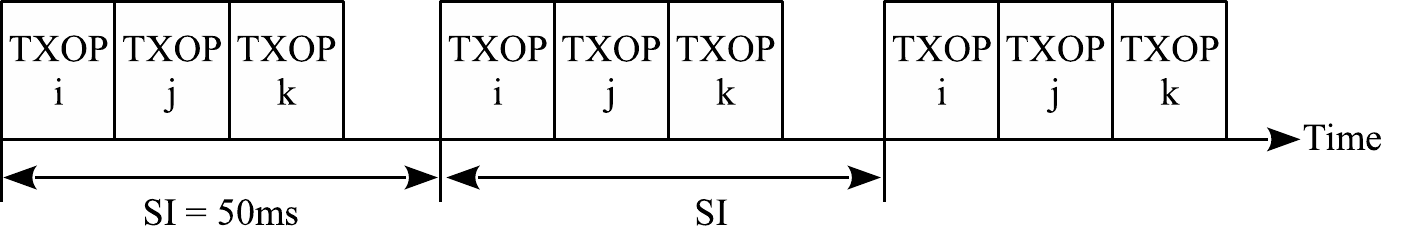}
		\caption{Schedule for streams from STAs i to k. The streams are scheduled in Round-Robin fashion govern by the admission control unit}
		\label{fig:ACU}
	\end{figure}
	
\end{enumerate}

\begin{figure}[!th]
	\centering
	\includegraphics[width=.6\linewidth]{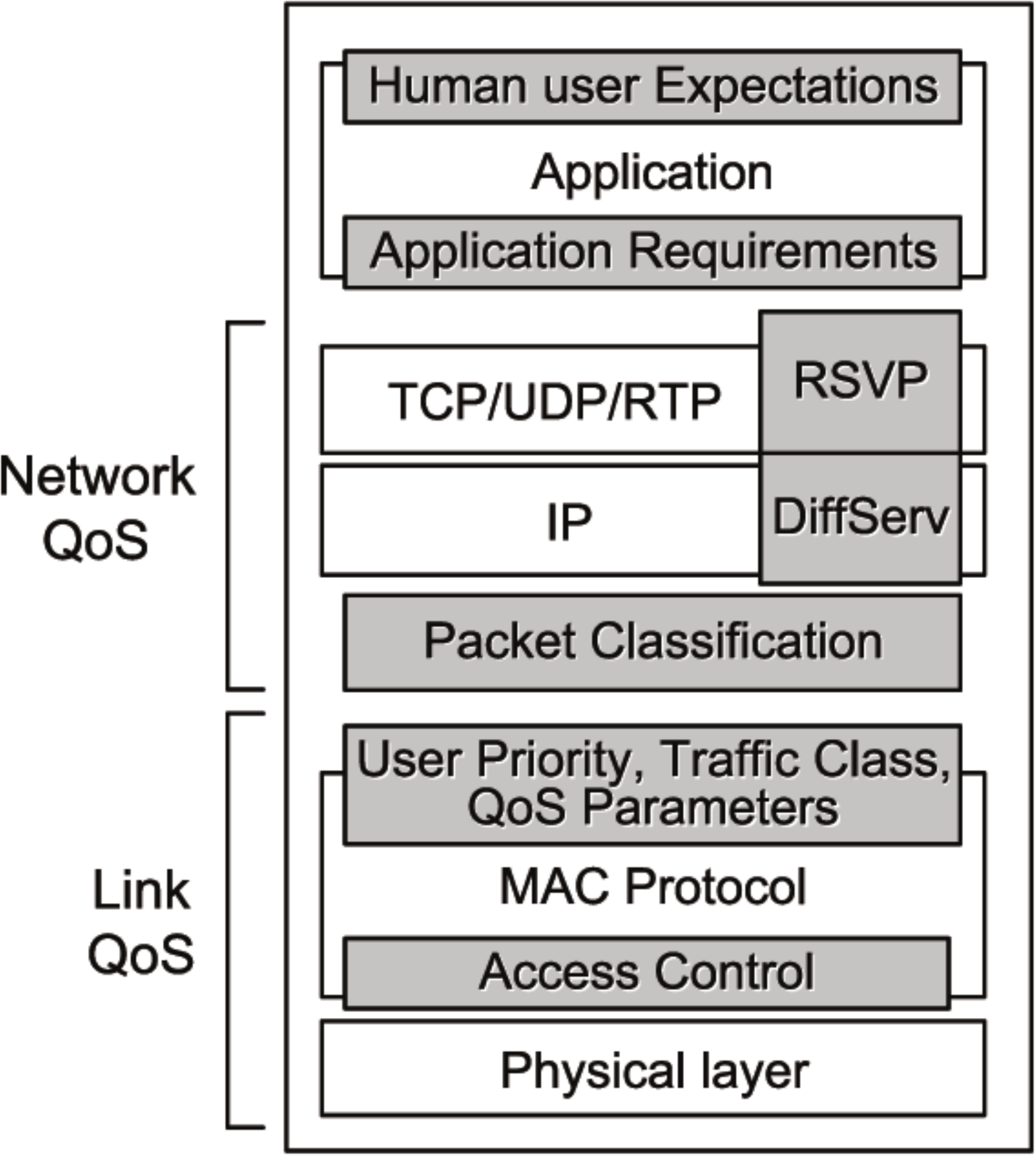}
	\caption{QoS architecture of the IP Network. The QoS parameters are defined in the MAC layer}
	\label{fig:IPQoS}
\end{figure}
\section{QoS challenges in IEEE802.11e WLANs}
\label{sec:QoSChalng}
QoS is the overall effect of the service performance, which defines the satisfaction degree of a service user and manifests itself in a number of subjective or objective parameters \cite{rec1994800}. There are two ways to investigate the QoS, subjective (perceptive) and objective (network) measurements. In the subjective measurement, the user involves to carry out a series of assessment tests, while in objective measurement, typical network performance throughput, packet loss, packet jitter and delay is evaluated. In order to meet the user satisfaction, the subjective QoS parameters shall be translated into a set of objective QoS parameters, e.g. throughput, delay and losses.

QoS could be supported in different ways at different protocol layers as illustrated in Fig.~\ref{fig:IPQoS}. Some applications have the capability to adapt the generated traffic to the conditions of the underlying network in order to meet user expectations. An example is the use of the \gls{RTP} and associated \gls{RTCP} \cite{jacobson2003rtp} to dynamically adapt the parameters of an audio and/or video streams, minimizing the losses due to congestion in the network \cite{busse1996dynamic}. Nevertheless, application layer mechanisms are usually not enough, since end-to-end QoS requires support in the lower layers of the protocol stack throughout the network nodes that the traffic must traverse from sender to receiver. However, this work mainly concerns with QoS provisioning at \gls{MAC} layers.

The QoS provisioning of diverse multimedia streams in a wireless environment imposes a chain of challenges due to many factors of \gls{OSI} model layers ~\cite{ISO7498_1994,Delsing2012,alani2014osi} ranging from traffic characteristics in application layer down to the wireless channels nature in physical layer. In this section, a review of the major challenges that may emerge when providing QoS for delay-sensitive applications in IEEE802.11e wireless networks.

\subsection{Adaptation to fluctuation of application profile}
\label{sec:Adpt2Aplica}

Generally, the application profile of a traffic is defined by the alternation of the traffic over the time. The QoS provision of a \gls{VBR} flow is  substantially influenced by the variation of the application profile over the time. The accurate estimation of the traffic at the application layer can significantly enhance the performance of underlying functions of \gls{MAC} layer to adapt its parameters according to these changes.

The \gls{VBR} video source can be generally classified into three main categories~\cite{Inan2006,Huang2009}: I) variable packet size with constant Generation Interval (GI), e.g., MPEG-4 videos; II) constant packet size with variable GI, e.g., Voice over Internet Protocol (VoIP); and III) variable packet size with variable GI, e.g., H.263.

The transmission of video streams can be significantly affected by the compression techniques used, such as MPEG-4 and H.263. The nature of the frame structure and the compression algorithm used along with the variations within video scenes can significantly influence the burstiness level of the stream ~\cite{trathgeb1993,hattacharyya2014}. The burstiness of a \gls{VBR} stream traffic increases the complexity of network resources management to ensure QoS support for continuous stream playback. Although, the reference design of the \gls{HCCA} scheduler is simple and efficient in supporting constant application profile, yet it is not adequate since it cannot address the fast-changing imposed by the \gls{VBR} bursty traffic, which hinders the performance of \gls{HCCA} by causing packets to wait for a longer time in their transmission queues.

In case of downlink traffic, from \gls{QAP} to \glspl{QSTA}, the \gls{QAP} is aware about its data queues and shall use its highest priority to seize the channel if it remains idle for a duration of \gls{PIFS} without undergoing back-off procedure. However, due to the fact that \gls{QAP} suffers from the lack of information about the uplink transmission queue status, an adaptive scheme is required to allow the scheduler to adjust its behavior based on the current application characteristics. Generally, adaptation to the application can be categorized according to its variability level-based in the three well-known types mentioned in \ref{sec:Adpt2Aplica}.

In \gls{MAC} layer, the uplink traffic profile can be determined using different ways, such as estimating the data buffer of the flow, predicting the packet generation time and/or traffic load at a specific time, or obtaining actual information through cross-layer architecture design. By having the traffic profile, the \gls{HCF} can adjust one or more of its functions such as polling \cite{ramos2007}, \gls{SI} assignment \cite{Grilo2003}, \gls{TXOP} allocation mechanisms \cite{Jansang2011} which allows it to instantaneously adapt to QoS requirement of the flow.

The QoS of VBR video transmission is ungoverned due to the fact that those packets are queued for a duration equivalent to SI until already-queued packets in the buffer are delivered.  Recall that during each SI, the reference HCCA scheduler allocates a fixed TXOP to each QSTA based on its mean rate requirements regardless the real VBR traffic changes. There are three QoS challenges relevant to Class I, II and III of VBR traffics.

\subsubsection{QoS Challenges of Class I video flows}
\label{sec:QoSClassIIChallanges}

HCCA scheduler fails to accommodate to variability Class I traffic which, in turn, leaves the wireless bandwidth in underutilization status. Assume, without loss of generality, that an identical TXOP duration is allocated for every QSTA, consequently, each QSTA will waste the same amount of unused TXOP ($T_u$). Thus, Equation \eqref{eq:ACU} can be rewritten as follows:
\begin{equation}
	\label{eq:ACUa}
	\frac{TXOP_{n+1}}{SI}+ \frac{TXOP-T_u}{SI} \leq \frac{T- T_{CP}}{T}
\end{equation}

According to Reference \cite{Qinglin2007}, using different SIs for different streams will improve the bandwidth utilization up to 50\%. In other words, the $T_u$ in Equation \eqref{eq:ACU} will be equal to $\frac{TXOP}{2}$. Therefore, the Equation \eqref{eq:ACU} can be again rewritten as follows:

\begin{equation}
	\label{eq:Bim}
	\frac{TXOP_{n+1}}{SI}+ \frac{TXOP}{2 \times SI} \leq \frac{T- T_{CP}}{T},
\end{equation}
which means that the number of admitted flows can be maximized to double the number of admitted flows when different SIs are used.

\subsubsection{QoS Challenges of Class II video flows}
\label{sec:QoSClassIChallanges}

In Class II, when $QSTA_{i}$, at any SI, exploits only portion of its allocated $TXOP_{i}$ at the traffic setup time, namely $T_{eff}^{i}$, leaving an unspent amount of $T_{u}^{i}$. Thus, the following relation can be held \cite{almaqri2016}:
\begin{equation}
	\begin{split}
		\sum_{i=1}^{N}{T}'_{i} & = T_{eff}^{1} + T_{eff}^{2} +\cdots + T_{eff}^{N}\\
		& = TD_{1} - T^{1}_{u} + TD_{2} - T^{2}_{u} +\cdots +TD_{N} - T^{N}_{u} \\
		& = \sum_{i=1}^{N} TD_{i} - \sum_{i=1}^{N}T^{i}_{u} \\
	\end{split} 
\end{equation}
where  $T^{i}_{u} \geq 0$, $\sum_{i=1}^{N}{T}D_{i}$ and $\sum_{i=1}^{N}{T}'_{i}$ is the total TXOP scheduled in any SI used in HCCA and ATXOP, respectively. It is worth noting that $TD_{i}$ is the TXOP duration of the $QSTA_i$ including  the poll overhead. Thus, the delay of $QSTA_{i}$ in an SI is computed as follows:
\begin{equation}
	D_{SI}^{i}=\sum_{j=1}^{i-1} (TD_{i} - T^{i}_{u}) + T_{L}^{i} + T_{poll} + 2 \times SIFS
	\label{eq:SIdlyAXTOP}
\end{equation} 

Altogether, the real QoS challenge is to minimize packet delay by minimizing the surplus amount, namely $T^{i}_{u}$.

\subsubsection{QoS Challenges of Class III video flows}
\label{sec:QoSClassIIIChallanges}

In video streams like H.263, the deviation comprises not only packet size but also shows up to high variation in generation interval which makes the matter much worse. In any \gls{SI}, scheduling a \gls{QSTA} based on its \gls{TSPEC} likely imposes allocating surplus of \gls{TXOP} duration which leads to wasting of the resources. This waste of resources due to the variations in data rate influences the efficiency of the scheduler that does not implement any recovery policy. Besides, due to the variation in the packet generation interval, perhaps there are some QSTAs that are not ready to transmit which will be considered as over-polling state. This waste of resources, due to the variations in data rate, influences the efficiency of the scheduler that does not implement any recovery policy. Overall, it hinders the meet of delay bounds requirements, which leads to a degradation in QoS provisioning. 

\begin{figure*}
	\centering
	\includegraphics[width=.8\linewidth]{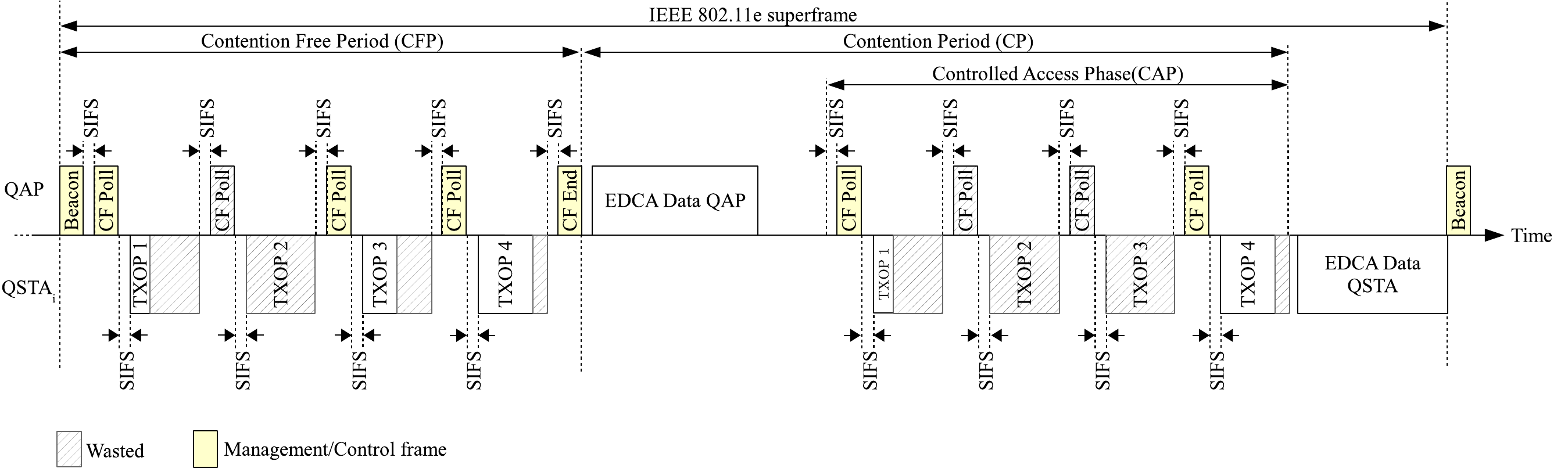}
	\caption{Wasting TXOP and poll issue with VBR traffic transmission}
	\label{fig:HCCAVideoTrans}
\end{figure*}

Consider the example illustrated in Figure \ref{fig:HCCAVideoTrans} where four QSTAs are polled for transmission in both \gls{CFP} and \gls{CAP}. In this example, $TXOP_{1}$, $TXOP_{2}$, $TXOP_{3}$ and $TXOP_{4}$ to $QSTA_{1}$, $QSTA_{2}$, $QSTA_{3}$ and $QSTA_{4}$, respectively. The wasted \gls{TXOP} and over-polling issues experienced using reference \gls{HCCA}, inspired from the example, are as illustrated in Figure \ref{fig:HCCAVideoTrans}.

\begin{itemize}
	
	\item \textbf{Over-polling of \glspl{QSTA}}	
	As illustrated in this example, due to the lack of awareness about the change in the traffic profile, some QSTAs may receive unwanted poll messages as their transmission queues are empty. $QSTA_{2}$ and $QSTA_{3}$ in \gls{CP}, and $QSTA_{2}$ in \gls{CFP} will respond with a null-frame causing unwanted delay to all QSTAs that may come after them in the same \gls{SI}.
	
	\item\textbf{Wasted \gls{TXOP} duration}	
	Since some $QSTAs$, such as $QSTA_{1}$, experience a high instant drops-down in data rate, only a short amount of the given \gls{TXOP} duration is utilized. In this case, the channel might remain idle for a period of time greater than the \gls{SIFS} and the control of the medium conveyed to \gls{AP} to poll the next station in the list. Even though the effect of wasted \gls{TXOP} duration in the packet delay is not as high as that caused by over-polling case, however, it is considerably can go high as the number of stations in the network increases.	
\end{itemize}

\subsection{Adaptation to varying network conditions}
\label{sec:Adapt2Net}
Due to the phenomena of path loss, multipath fading, shadowing, and interference, wireless networks likely suffer from \gls{SINR} \cite{Grilo2003}. The fluctuation of the underlying channel capacity will hinder the QoS provisioning for time-sensitive applications. Consequently, two possible ways to be applied on the QoS algorithms in order to encounter this challenge and meet the required QoS needs. The first one is by computing the transmission time for the packets based on the minimum physical bit rate announced. By doing so, the QoS is guaranteed, however, this technique gives rise to degradation in bandwidth efficiency as the bandwidth might get higher anytime while only the minimum link rate is considered. The second one is to encourage the QoS algorithm to take into account the link adaptation mechanism of WLANs over the time.

Although, piggybacking feature of \gls{HCCA} is basically designed to improve the channel capacity, it may inversely behave when a station experience successive retransmission or channel noise. This issue has been referred to in \cite{Lee2007} as "the piggyback problem as the low physical transmission rate". If any \gls{QSTA} was transmitting at a low physical rate due to channel error, \gls{QAP} will accordingly decrease the transmission rate of the piggybacked \gls{CF-Poll} frame. This, in turn, will result in channel efficiency degradation and will increase the TSs' frame delay of other stations involving in the \gls{NAV} process.

In case of \gls{VBR} traffic transmission over WLANs  and apart from the issues and challenges of \gls{HCCA} reported in \cite{Inan2006,madhar2012}, the major issue of the reference scheduler is the unawareness about the inherent wireless time-varying channel condition\cite{arora2010}. Keeping aware about the channel status has a major impact on the scheduler performance as it can potentially degrade the service differentiation process, even though, \gls{HCCA} has been observed to perform well in heavy loaded network \cite{cicconetti2005,Chen2011}, especially with the emergence of several physical layer technologies such as \gls{AMC} schemes.

\subsection{Bandwidth utilization}
\label{sec:BwUtil}
In \gls{HCCA}, after receiving an ADDTS-request from the station, the scheduler needs first to calculate the required \gls{TXOP} duration taking into account its \gls{TSPEC} parameters. Thereafter, the used admission control mechanism will check the ability to accept the new \gls{TS}. If the new \gls{TS} is accepted, the \gls{SI} will be computed as the minimum among all delay bounds of admitted streams, which is enough to meet the most urgent delay requirement to guarantee the required QoS service for the admitted streams. Finally, the round robin approach is used to allocate TXOPs to the involved station. Even though the use of this design is very simple and straightforward, it still suffers from some challenging issues related to the efficient use of the bandwidth. Indeed, the use of round robin approach in \gls{HCCA} scheduler to serve all TSs in one \gls{SI} might lead to over-allocating the bandwidth, which in turn leads to under-utilizing the channel bandwidth. Moreover, the waste of the wireless bandwidth may reach up to more than 50\% in some cases \cite{zhao2007}.

In fact, based on the minimum physical rate and the characteristics of the incoming TSs, the \gls{ACU} decides the number of admitted TSs to which the wireless resource will be allocated. This approach leads to allocating a constant amount of resource to every \gls{TS} using the mean of single physical transmission rate, which is not compatible with the condition of current wireless bandwidth, especially \gls{VBR} traffic. In other words, the \gls{ACU} should consider both the physical layer and the service specific QoS parameters in order to be able to achieve effective bandwidth utilization \cite{lee2010}.

With noticeably \gls{VBR} flows, one of two scenarios likely occurs at some specific SIs. In the first scenario, the data rate becomes lower than the average value determined in the \gls{TSPEC}, thus, the allocated \gls{TXOP} will not be completely consumed which is considered as a wasting of resources. As for the second scenario, the data rate becomes greater than the average value determined in the \gls{TSPEC}, thus, the assigned \gls{TXOP} will not be enough to transmit the relevant data which increases the end-to-end delay of the flow. The possible solutions to solve these two problems as explained in these references \cite{cecchettielAL2012,ruscelli2014} are: (1) By increasing the \gls{TXOP} duration to the average \gls{TXOP} of traffic for the first case, knowing that it will reduce the bandwidth utilization, especially if the data rate is dropped down. (2) By applying the bandwidth reclaiming approach \cite{Cecchetti2012,lo2007,rashid2008,ruscelli2011}.

\subsection{Network resources management}
\label{sec:NetResMan}
Indeed the \gls{HCF} of IEEE802.11e protocol is targeted to the provisioning of QoS throughout the service differentiation, yet the proper network resource management, such as coordinating between distributed (CP) and controlled (CAP) periods and link layer resources still in request \cite{ramos2005}. In addition, a feasible \gls{ACU} scheme is also required, which in such way can ensure that the QoS requirements are satisfied.
The \gls{HCCA} scheduler operates based on the static configuration of its traffic \gls{TSPEC} parameters where they are constantly served for their lifetime to enforce resource sharing with ensuring that the desired QoS constraints are met. To this aim, a good resource utilization is often left to the heuristic network administrator know-how. However, this constant resource sharing policy might highly cause a scarce bandwidth utilization since it cannot adapt to the transformation of the traffic profile and the lifetime due to dynamic \gls{VBR} traffic evolution.

As a resolution to this issue, a bandwidth sharing strategy is suggested to rely on a criteria which is driven by the performance \cite{navarro2010}, in which a common performance metric is recommended to be defined to differentiate between the traffic streams based on their performance requirements.
\section{Classification of QoS support for multimedia traffic approaches in IEEE802.11e WLAN}
\label{sec:classification}
In general, the enhancement approaches in IEEE802.11e protocol can be classified based on the access medium control fashion into distributed control and centralized control enhancements. In IEEE802.11e, \gls{EDCA} operates based on the distributed access control while the \gls{HCCA} represents the centralized access control. In \cite{ni2004}, many QoS enhancements for 802.11 WLAN have been proposed and classified along with their advantages and disadvantages. Another survey in \cite{luo2011} has focused on the QoS provisioning in both \gls{EDCA} and \gls{HCCA} over IEEE802.11e networks. The \gls{HCCA} enhancement approaches can be themselves classified into different categories according to several aspects such as the functional, structural, environmental and location aspects. In \cite{gao2005} and \cite{liu2009}, the authors presented a survey of various admission control in IEEE802.11e and they classify schemes based on several aspects such as Measurement-Based, Model-Based and Hybrid schemes. In \cite{piro2012}, the delay-EDD based scheduler has been compared to the feedback control based scheduler in order to provide a better comprehension about the so-called packet scheduling in 802.11 WLANs. Below is a short description of the different possible ways of IEEE802.11e approaches classification.

\begin{itemize}
	\item \textbf{Traffic flow direction:} In infrastructure mode of IEEE802.11 WLANs, the traffic directions would be either downlink and uplink. "Downlink" refers to a traffic flow transmitted from \gls{AP} to a mobile device, while "uplink" refers to a flow with a reverse direction. IEEE802.11e enhancements can be tailored to enhance the performance for either downlink or uplink traffic or in some cases for both directions.
	
	\item \textbf{Targeted environment:} Although IEEE802.11e \gls{MAC} was originally designed for wireless infrastructure networks and widely used in WLANs, there have been some enhancements for adapting IEEE802.11e to work with other networks such as the improvement of polling and scheduling scheme over IEEE802.11a/e \cite{Schaar2006}, IEEE802.11p networks \cite{zhang2013}, Ad-hoc Wireless Networks in \cite{Fiandrotti2008} or in Integrated model of IEEE802.11e and IEEE802.16 \cite{Naeini2012,Lee2009wimax}.
	
	\item \textbf{Delay-EDD based and feedback control based:} 
	
	The pre-knowledge of packets arrival time is only possible for the downlink. While in the uplink, neither the delays of the head of line packets nor the quota of bandwidth needed by each flow are possible to be known by the access point. For this reason, IEEE802.11e schedulers have been categorized into the earliest due date and the feedback control class. A thorough comparison between these types has been presented in \cite{piro2012}.
	
	\item \textbf{Layered vs. cross layer:} IEEE802.11e enhancements can be introduced in two structures, cross-layer and layered approaches. The cross-layer approaches rely on interactions between two layers of the \gls{OSI} architecture.  These approaches were motivated by the fact that providing lower or higher layer information to \gls{MAC} layer to perform better. The layered approaches rely on adapting \gls{OSI} layers independently of the other layers. Cross-layer is a promising direction to improve the overall performance of WLAN since it takes into account the interactions among layers \cite{shankar2007}. Thus, several enhancements \cite{Cicconetti2007,noh2010,luo2011cross} prefer to use cross-layer design for obtaining accurate information for scheduling purposes.
	
	\item \textbf{Technique or mechanism used:} The \gls{HCCA} scheduling approaches can be classified based on the techniques and/or mechanisms used in the design. In the literature, a diverse techniques were developed for \gls{HCCA} scheduling to boost its performance for multimedia transmission over error-prone WLANs such as estimation based approaches \cite{Grilo2003,Ansel2004}, predicting traffic profile \cite{Yoo2002,kuo2011}. Moreover, some of these approaches modified one or more of \gls{HCCA} mechanisms such as \gls{TXOP} assignments \cite{ju2013,lee2013,noh2011,Huang2008}, polling mechanism \cite{Chen2004,Ramos2012,Son2004} or \gls{ACU} \cite{noh2010,Kim2006,Zhu2007,lee2010}.
	
	\item \textbf{Analysis method used:} The approach might be analyzed and/or evaluated using one of three methods, namely analytical model \cite{Harsha2006,Cecchetti2008,leonovich2013,Jansang2013}; simulation experiment \cite{ng2013,ruscelli2012,Kim2006TWOPoll} and test bed \cite{Ng2005}. It is worth noting that the analytical model usually is done to capture the characteristics and the shortcoming of the approach and prior to the proposal of a solution. Although, in simulation and test bed methods used for evaluating the proposed scheme, they might be carried out to provide a preliminary study to investigate a particular issue in the existing scheme for possible remedy.
\end{itemize}

\section{QoS Enhancements in IEEE802.11e controlled access mode}
\label{sec:HCCA_enhancement}
This section presents some of the leading approaches proposed in the literature to improve the QoS provisioning for multimedia traffic. More emphasis has been put on the transmission of \gls{VBR} video streams in IEEE802.11e WLAN. The approaches are classified into six sets based on the strategy used to improve \gls{HCCA} performance. However, some approaches can be matched to several types of strategies, but are only classified to their main strategy. Moreover, in the layered approaches the focus was only on the enhancements of \gls{HCCA} at the \gls{MAC} layer. The representative approaches are defined and their mechanisms are described along with a discussion concerning their strengths and weaknesses in improving QoS performance in IEEE802.11e WLAN. In addition, a comparison of the main characteristics of various \gls{HCCA} approaches is provided for each category. Besides some mathematical models that study and provide insights to improve the \gls{HCF} functions have been presented which can provide a promising avenue for further research and investigation.

\subsection{HCCA polling enhancements}
\label{sec:pollEnh}
The polling mechanism of the legacy \gls{HCCA} is responsible for the scheduling and the allocation of \gls{TS} based on their fixed reservations. Thus, the efficiency of this mechanism highly depends on the accuracy of the flow specification declared to the \gls{HC}. Yet, as the flow profile of \gls{VBR} might highly vary over the time, the allocation based on fixed reservations will cause degradation in quality of multimedia flows even when the channel resource is surplus. More particularly, several issues may affect the efficiency of the \gls{HCCA} such as the inefficient Round-Robin scheduling algorithm, the overhead induced by the poll frames, and the lack of coordination between the APs of the neighboring \gls{BSS}. The representative approaches that address these issues and even more are summarized as follows.

\begin{figure}[!h]
	\centering
	\includegraphics[width=0.8\linewidth]{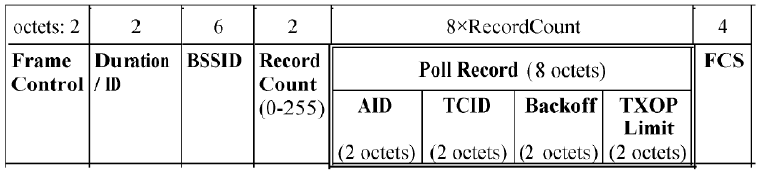}
	\caption{CP-Multipoll frame format}
	\label{fig:CP-Multipolling}
\end{figure}

\textbf{CP-Multipoll} is a robust multipolling mechanism aims to increase the channel utilization and to minimize the corresponding implementation overhead, which can be robust in  error-prone environments like WLANs \cite{Lo2003}. Moreover, the proposed scheme provides a polling schedule to ensure the bounded delay requirements of real-time traffic and it also provides an admission control mechanism. The main aim of this scheme is to design an efficient polling mechanism, due to its high impact on the performance of \gls{HCCA}, which is able to serve both CBR and VBR real-time traffic. Unlike SinglePoll schemes where every \gls{STA} receives one poll frame when polled, CP-Multipoll aggregates many polls in a single multipolling frame incorporating the \gls{DCF} into the polling mechanism. The frame format of CP-Multipoll scheme is as shown in Fig.~\ref{fig:CP-Multipolling}.

Basically, CP-Multipoll conveys the polling order into the contending order. This can be achieved by assigning different back-off values to the streams in the polling list with accordance to their ascending order in the polling list and allow the back-off to execute as soon as they receive the CP-Multipolling frame. Besides minimizing the polling overhead by transmitting one polling message for all \glspl{QSTA} in the polling list instead of sending polls as many \glspl{STA}, the proposed scheme has other advantages over other multipoll schemes. The bursty traffic is better supported since the \gls{STA} holds the channel only for a period needed to transmit its local buffered data. Moreover, in \gls{DCF} access mode, if the \gls{STA} does not use the poll frame due to empty data buffer, the other STAs in this polling group will immediately detect that the channel is idle and it will advance the starting of channel contention.
However, the proposed mechanism is prone to hidden terminal problems since each \gls{STA} will decrement back-off counter when it senses that the channel is idle. Thus, if hidden terminals exist in the network, different STAs will complete their back-off simultaneously and collision will happen. Due to the inherent hidden node issue of infrastructure wireless networks, CP-Multipoll cannot guarantee that all STAs in the \gls{BSS} can sense the transmission of other STAs. In this case, the station will transmit its data immediately upon the expiration of its back-off timer leading to a collision.

\begin{figure}
	\centering
	\includegraphics[width=0.9\linewidth]{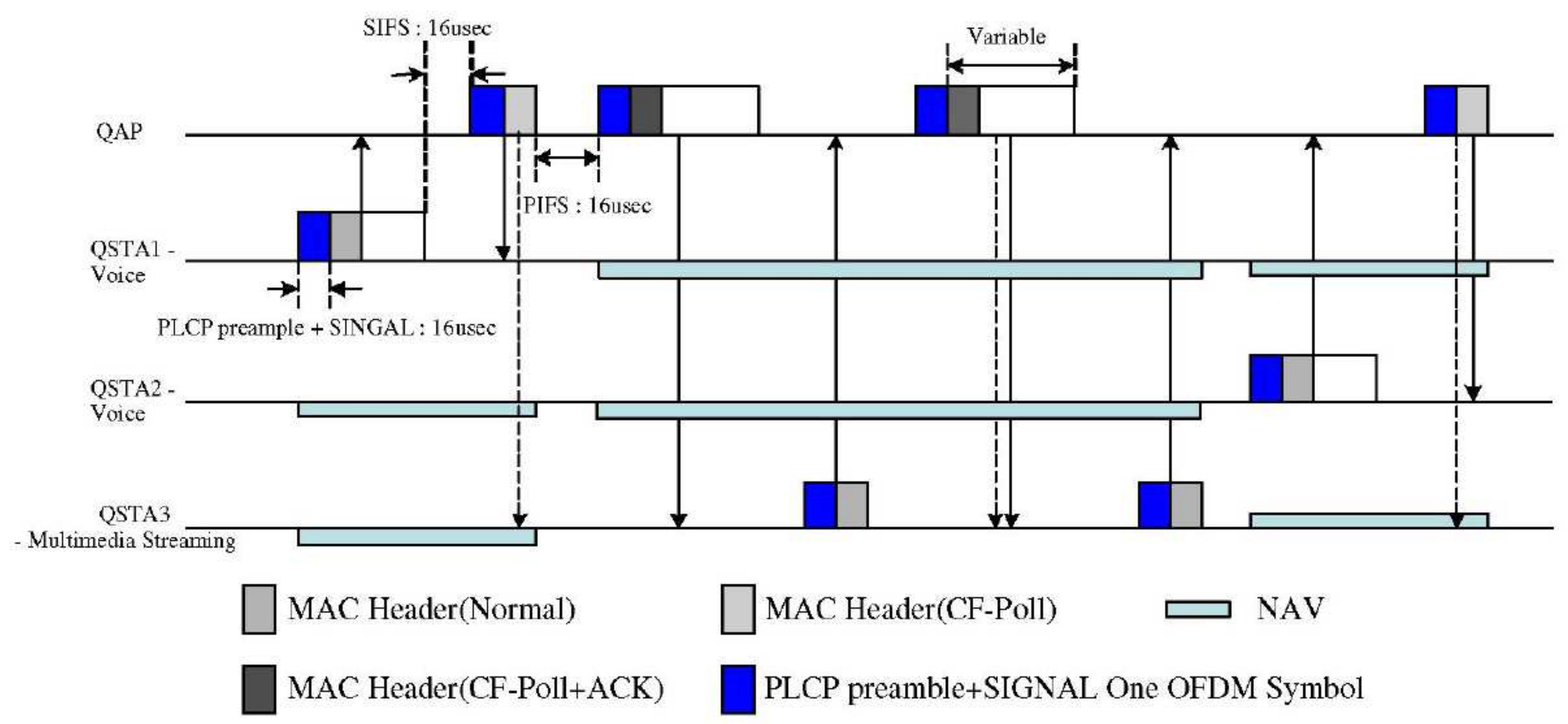}
	\caption{CF-Poll piggyback issue with an example of piggybacking CF-Poll on data frame}
	\label{fig:CFPoll}
\end{figure}

\textbf{CF-Poll piggyback scheme} is presented by Lee and Kim \cite{lee2006} to optimize the usage rule of the CF-Poll piggyback scheme as defined in the IEEE802.11 standard \cite{IEEEStand1999} according to the \gls{TS} load and the minimum physical transmission rate of a \gls{QSTA} which suffer the deep channel fading. Consider the case of piggybacking, the CF-Poll in the QoS-ACK frame from \gls{QAP} to $QSTA_{3}$, illustrated in Fig.~\ref{fig:CFPoll}, must be listened by all \gls{QSTA} in the \gls{BSS}. If any of the QSTAs experience low physical rate, which implies that $QSTA_{3}$ requires more time to receive the frame, the delay for all other QSTAs will be increased and the channel efficiency will be decreased.

Motivated from the aforementioned issue, the proposed work provides a guideline for the optimal usage of the CF-Poll piggyback scheme in IEEE802.11e and IEEE802.11n protocols. Simulation-based results reveal that the frame transmission delay is majorly affected by the minimum physical rate when CF-Poll is piggybacked in the QoS data frame while it is slightly influenced by the traffic load. The results show an inverse relationship between the \gls{CF-Poll} piggyback scheme and the traffic load. Despite the presented analysis and guidelines, the recommendations reckon on a number of assumptions that are: the traffic is \gls{CBR} and each \gls{QSTA} has only one \gls{TS} calculated based on the Equation~\eqref{eq:N} which cannot be suitable for supporting the transmission of multimedia applications with variable profile.\\

\begin{figure}
	\centering
	\includegraphics[width=0.8\linewidth]{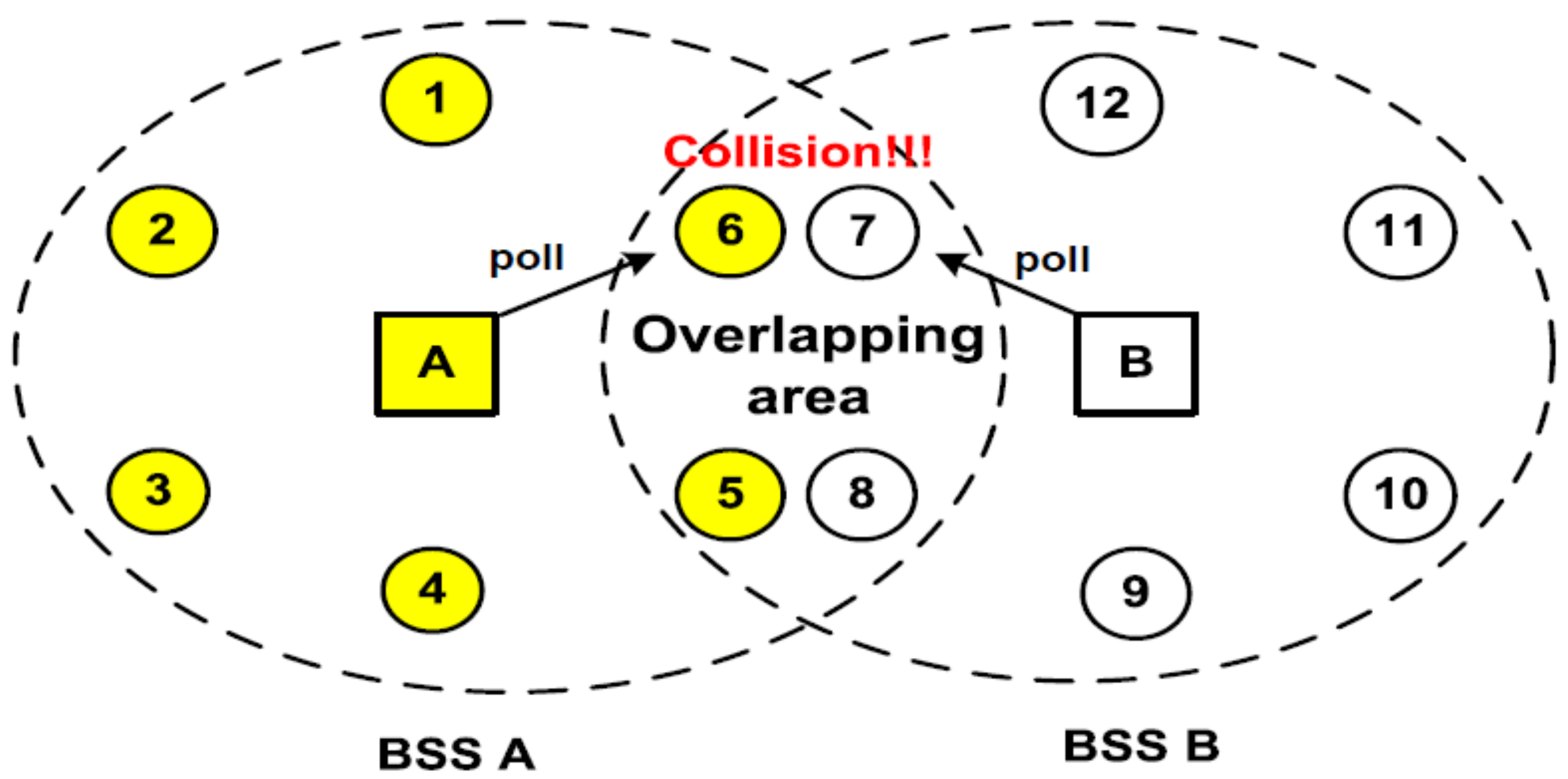}
	\caption{Collision due to polling the STAs in the overlapping area}
	\label{fig:DEBScheme}
\end{figure}

\textbf{\gls{DEB} method for \gls{HCCA}} is an enhancement of \gls{HCCA} which performs virtual polling through sensing the carrier of the wireless channel \cite{Huang2010}. This technique highlighted the issue of the collision incurred due to polling the nodes in the overlapping area of two adjacent BSSs at the same time. This actually occurs due to the lack of coordination in \gls{HCCA} between the adjacent APs. Consider the nodes 5, 6, 7, 8 in the overlapping area illustrated in Fig.~\ref{fig:DEBScheme}. Since \gls{AP} A cannot hear \gls{AP} B, therefore the collision occurs between the nodes in the overlapped area.
\gls{DEB} uses a similar idea of sensing the carrier of \gls{EDCA} since it manifests high robustness and flexibility controlling the medium at the overlapping BSSs. A virtual polling has been achieved in a distributed manner. The \gls{DEB} arranges the back-off timer of station to guarantee that the polled stations will have different back-off. When the back-off timer expires, the station can be polled without colliding with others. However, \gls{DEB} is only functioning in \gls{CFP} whereas \gls{HCCA} is supposed to work in both \gls{CFP} and \gls{CP}, for this reason one of the significant merits of \gls{HCCA} will be untapped. Moreover, there is no clear consideration of the readiness of the station, \gls{STA} with no data ready to send will be given a \gls{TXOP} which, in turn, be wasted.

\textbf{\gls{NPHCCA}} is presented in \cite{Chen2011} to provide an enhancement over \gls{HCCA} mechanism. Since the \gls{VBR} traffic exhibits variability in packet generation time, the station will not always have pending data to transmit, thus, it will waste time for the \gls{AP} to send polling messages to the stations that have no data to transmit. For this reason, the proposed solution modifies the \gls{HCCA} scheme in such way it allows stations that have pending frames to report their readiness status to \gls{AP} through exchanging messages. Then, the \gls{AP} schedules the only ready stations in appropriate transmission sequence.

\begin{figure}
	\centering
	\includegraphics[width=0.8\linewidth]{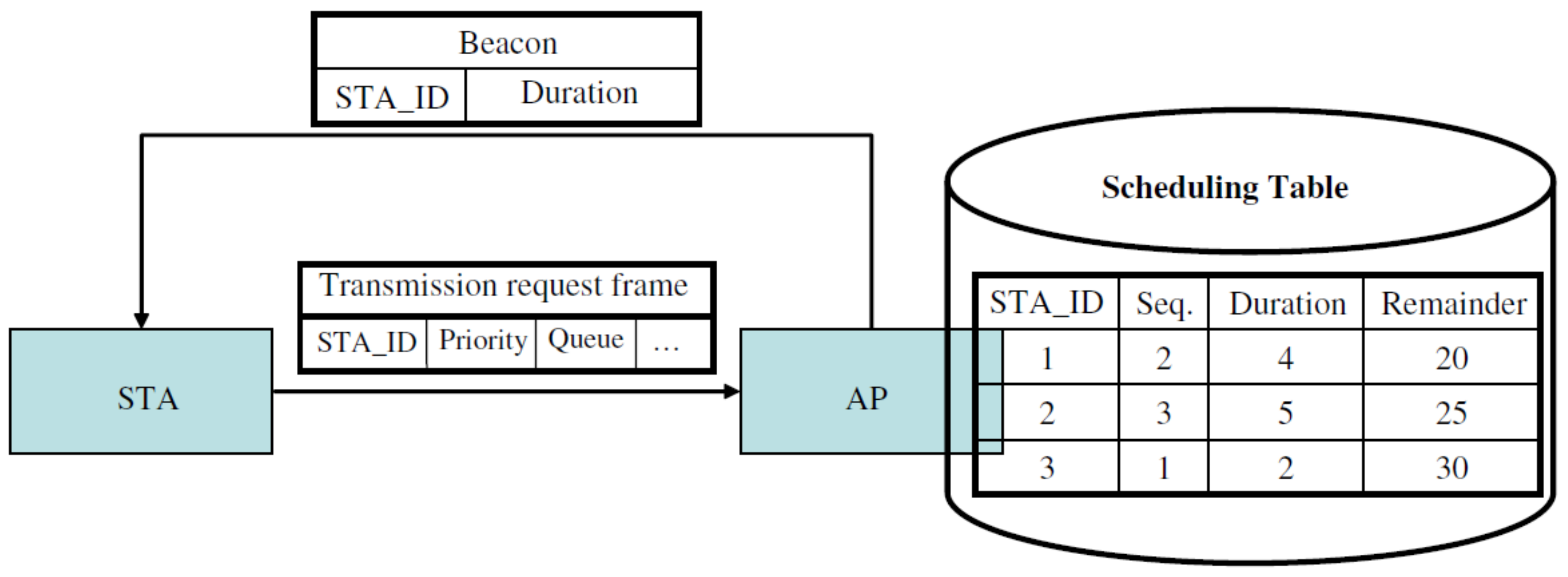}
	\caption{\gls{NPHCCA} mechanism}
	\label{fig:NPHCCA}
\end{figure}

The mechanism of the \gls{NPHCCA} is carried out throughout a sequence of messages exchanging. First, a station with data will send a transmission frame request to the \gls{AP} in order to update it about its transmission queue status, including information such as required Priority, Queue status, etc. A station only sends this frame after it receives the beacon message from the \gls{AP} and senses whether the medium is idle for \gls{SIFS}. Accordingly, the \gls{AP} maintains this information in its scheduling table. Finally, the \gls{AP} determines a transmission sequence and notifies stations to transmit data according to this transmission sequence broadcast in the beacon messages. Fig.~\ref{fig:NPHCCA} demonstrates The components of the \gls{NPHCCA}.
Although, \gls{NPHCCA} has shown improvement in the transmission delay when the network is light-loaded, the performance was similar to that of \gls{HCCA} when the network is heavy-loaded. Besides, the messaging exchange of the beacon and transmission request frames added extra overhead to the network, especially when the number of the nodes increases.

\begin{figure}[!h]
	\centering
	\includegraphics[width=0.8\linewidth]{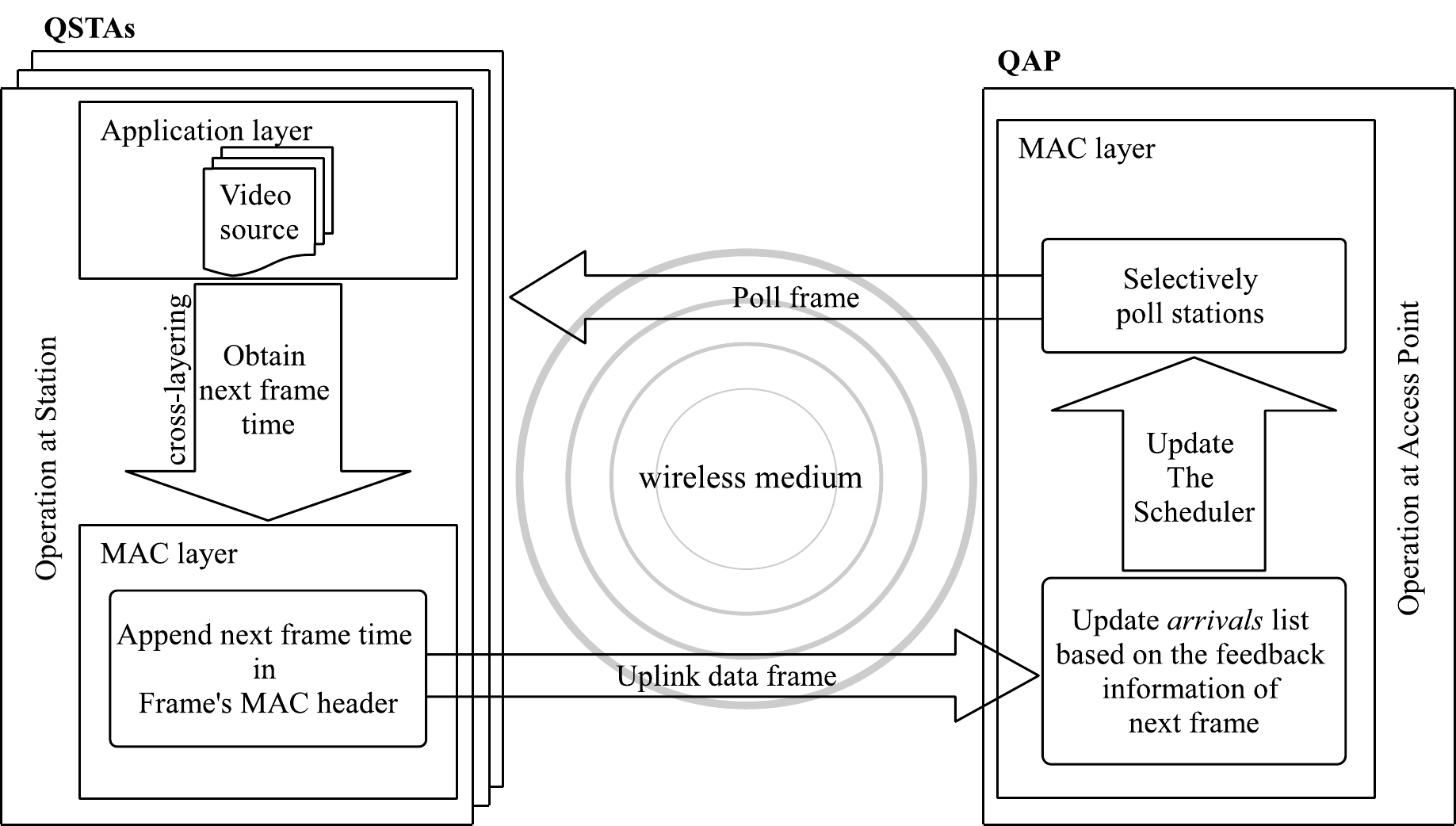}
	\caption{F-Poll scheme mechanism}
	\label{fig:FPollMechanism}
\end{figure}

\textbf{F-Poll} In Feasible Polling Scheme (F-Poll) \cite{maqri2015}, the application layer gives the accurate arrival-time of the upcoming data frame over the uplink connection to the MAC layer, where this approach is known as a cross-layering approach. F-Poll is suitable for both type II and III of video types categorized in Subsection~\ref{sec:Adpt2Aplica}, where the exact information of the next inter-arrival time is sent to the QAP in order to enhance the scheduling of the TSs. In order to avoid polling a station that have no ready data to transmit, a decision is made of whether to poll the relevant station in the upcoming SI or not directly after receiving a data frame. As a result, the packet access delay is minimized and a great amount of unused TXOP duration is conserved which efficiently enhances the channel utilization. Fig.~\ref{fig:FPollMechanism} elaborates the F-Poll Mechanism.

\textbf{AMTXOP} \cite{almaqri2016} like D-TXOP \cite{Almaqri2013}, the \gls{AMTXOP} calculates the TXOP for a certain data stream based on the actual frame size. Since the polling messages can increase the overheard among all QSTAs, the BSS broadcasts one multi-polling message to the QSTAs in a single SI instead of sending one polling message for each. This approach minimizes the polling overhead and also reduces the packet delay which significantly improves the bandwidth utilization. Due to this integration, the AMTXOP outperforms both HCCA and its ancestor, D-TXOP, in terms of channel utilization and packet delay.

\subsection{\gls{TXOP} allocation enhancements}
\label{sec:TXOPEnh}
Usually, if a QSTA's buffer queue is empty during its \gls{TXOP} because of a non-uniform data flow from the upper layer, the media will be unutilized for the whole \gls{TXOP} of the station. However, according to the 802.11e standard, the \gls{QSTA} should send a QoS-NULL frame to the \gls{QAP} to enforce it to start polling other sessions immediately \cite{Jansang2013}. On the other hand, if the allocated \gls{TXOP} is not enough to send the backlogged packets, these data will be served in the next \gls{SI} causing more delay and might impair the designated QoS requirements \cite{rashid2008}. Several techniques have been presented in the literature to address the limitation of the \gls{TXOP} assignment mechanism of IEEE802.11e, we overview here some representative approaches.

\textbf{Scheduling Based on Estimated Transmission Times-Earliest Due Date (SETT-EDD)} \cite{Grilo2003} has proposed a novel scheduling technique for the so-called IEEE802.11e \gls{HCF}. A simple mechanism similar to the \gls{CAP} timer has been employed to limit the polling-based transmission in \gls{HCF} which so-called \gls{TXOP} timer. This \gls{TXOP} timer increases at a constant rate equal to the proportion of that \gls{TXOP} duration to the minimum service interval ($TD/mSI$), which reflects the fraction of time consumed by the station in polled TXOPs. The maximum value of this timer is equal to the maximum \gls{TXOP} duration ($MTD$). The consumed time by a station in a polled \gls{TXOP} is subtracted from the \gls{TXOP} timer by the end of the \gls{TXOP}. Thus, the station can be polled only if the \gls{TXOP} timer value is greater than or equal to the minimum \gls{TXOP} duration ($mTD$), which guarantees the transmission of at least one data frame at the minimum \gls{PHY} rate.

Since the \gls{TXOP} is allocated in SETT-EDD based on earliest deadlines, the transmission delay and data loss have been reduced. That is why SETT-EDD shows flexibility and considered a representative dynamic scheduler, as well as it provides compatibility to the link adaptation implemented in the commercial WLANs. However, it still lacks an efficient technique to be able to calculate the accurate required \gls{TXOP} for every \gls{QSTA} transmission instead of estimating \gls{TXOP} based on the average data rate of each \gls{TS} and the packet time interval between two consecutive transmissions.

\textbf{Adaptive Resource Reservation Over WLANs (ARROW)} is another algorithm where the \gls{TXOP} assignment is calculated dynamically based on the queued data size of the QSTAs \cite{Skyrianoglou2006,Passas2006}. In ARROW, the SETT-EDD \cite{Grilo2003} has been extended, where the available bandwidth is allocated based on the existing amount of data which is ready for transmission in every \gls{STA}. In contrast to SETT-EDD, which allocates the channel bandwidth based on the expected arriving data in every \gls{STA}. In this mechanism, \gls{QSTA} advertises the size of the total queued packets waiting for transmission with every poll. This information is piggybacked with the data frame prior the sending back to \gls{QAP}. So, the next \gls{TXOP} allocation for any particular stream will be calculated based on the advertised queue size.
\begin{figure}
	\centering
	\includegraphics[width=0.8\linewidth]{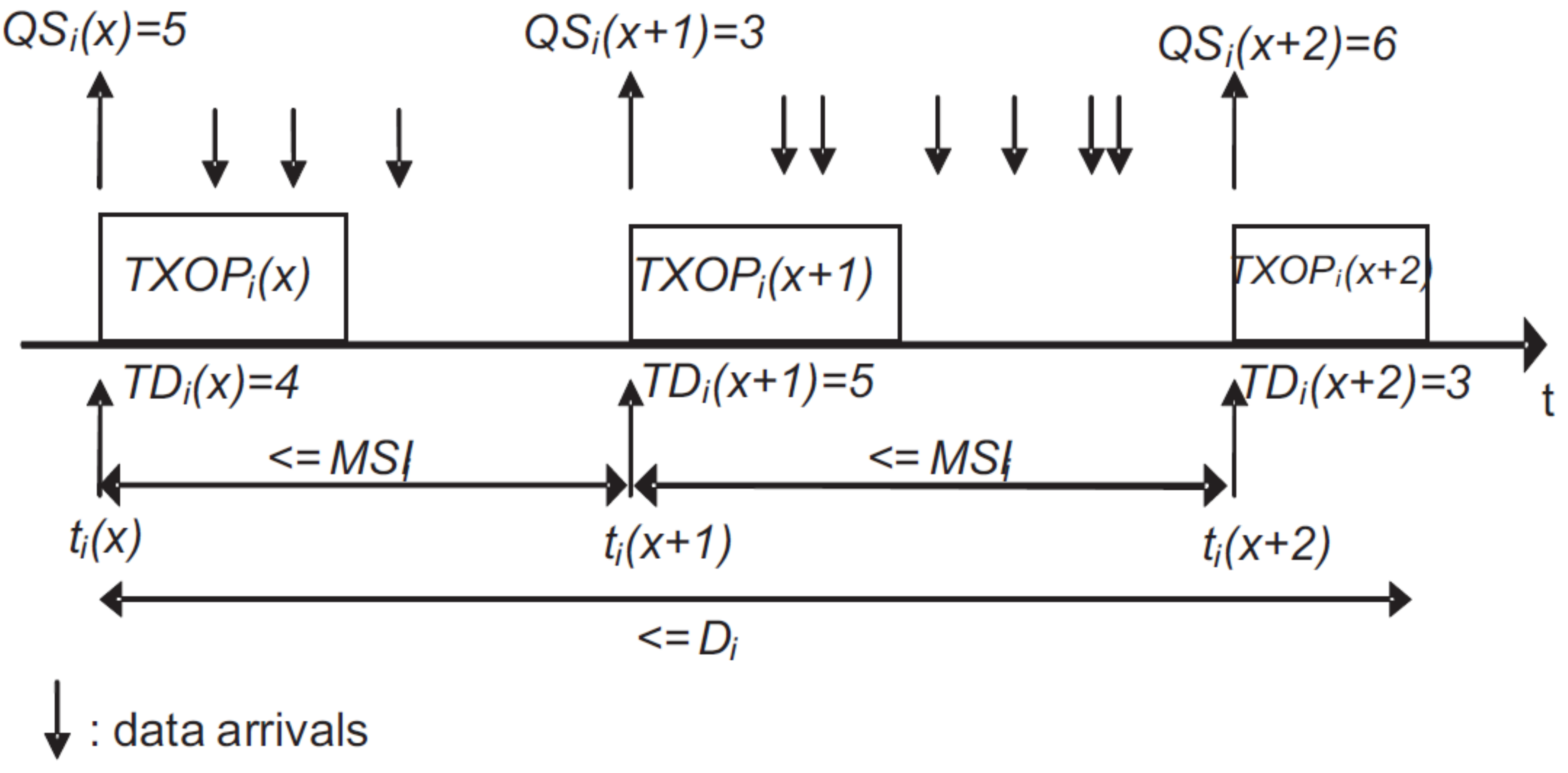}
	\caption{The ARROW Mechanism}
	\label{fig:ARROW}
\end{figure}
In this algorithm, the channel is allocated based on the exact transmission requirements for each \gls{QSTA}, which is expressed by the \gls{QS} field indicated during the previous \gls{TXOP}. By doing so, the \gls{TXOP} is assigned to meet the transmission requirement at the time when the previous \gls{TXOP} assignment is made and consequently the data buffered in the \gls{QSTA} is taken into account at any \gls{SI} leading to efficient adaptation of bandwidth allocation to actual requirements. Specifically, as illustrated in Fig.~\ref{fig:ARROW}, data arrive during $[t_{i}(x), t_{i}(x + 1)]$ can only be transmitted after the elapsing of $t_{i}(x + 2)$, which results in a delay of packets for at least one \gls{SI}.

\begin{figure}
	\centering
	\includegraphics[width=.8\linewidth]{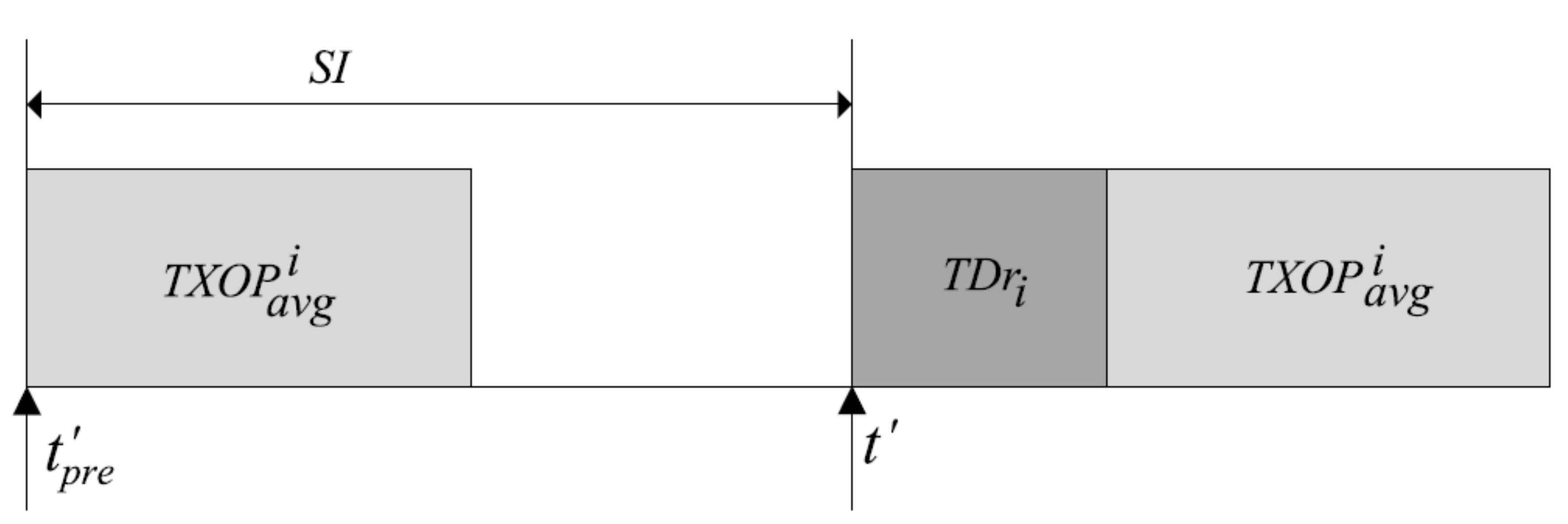}
	\caption{TXOP assignment in enhanced EDD}
	\label{fig:EnhancedEDD}
\end{figure}

\textbf{Enhanced \gls{EDD} QoS scheduler:} presented by \cite{Lee2009} and it is an EDD-based algorithm mainly aims at addressing the above-mentioned weakness of the ARROW scheduler. Similar to ARROW scheduler, the Enhanced \gls{EDD} also uses the queue length information like ARROW. However, the Enhanced \gls{EDD} estimates the number of arriving packets immediately after the end of the previous transmission, as shown in Fig.~\ref{fig:EnhancedEDD}. Thereafter, it calculates just the enough \gls{TXOP} to clear up the buffer queue by the end of current transmission. To reduce the average delay, when the buffer is not empty after the current transmission completes, the next \gls{SI} begins earlier, which can be achieved by changing the value of mSI and \gls{MSI}.
The \gls{TXOP} allocation in Enhanced \gls{EDD} is calculated for each station $STA_{i}$ as the summation \gls{TXOP} calculated exactly as in ARROW and a duration enough to transmit the packets generated during the current \gls{SI} as below
\begin{equation}
TXOP_{i}=TXOP_{avg}^i+TDr_{i}
\label{eq:EnhancedEDD1}
\end{equation}
where $TXOP_{avg}$ is calculated as follows
and $N_{curSI}$ is the expected number of packets generated from time $t_{pre}$ until $t'$.

\textbf{Dynamic \gls{TXOP} \gls{HCCA} \gls{DTH}} involves a bandwidth reclaiming mechanism into a centralized \gls{HCCA} scheduler in order to improve the transmission capacity and to provide additional resources to \gls{VBR} TSs \cite{cecchettielAL2012}. The main concept of \gls{DTH} is to prevent wasting the underutilized portion of transmission time in order to allocate it to the next polled station that needs longer transmission period. This approach relies on the unspent amount of the \gls{TXOP} from the previous poll time of a $QSTA_{i}$ as follows
\begin{equation}
TXOP_{i}= \left\{\begin{matrix}
TXOP_{AC(i)} & \text{if } T_{spare}\equiv 0 \\
t_{est(i)+T_{spare}}& \text{if } T_{spare} > 0
\end{matrix}\right.
\end{equation}
If there is no surplus \gls{TXOP} duration from previous poll time, which implies that the station exhausts the whole \gls{TXOP} duration. The next \gls{TXOP} duration will be the same as the one calculated in Equation~\eqref{eq:TXOP}. Otherwise, it will be calculated as the summation of the unused \gls{TXOP} duration and the estimated transmission time, computed through the \gls{SMA} of the effectively utilized duration in the previous polling intervals. Simulation results show that this approach can improve the performance, especially in terms of transmission queue size, data loss and delay, and the approach can absorb and follow the variation of VBR. Additionally, another analytical study confirms that the \gls{DTH} approach has no effect on the policy of the centralized scheduler. However, the estimation of transmission time using Moving Average needs more investigation as the \gls{VBR} traffic tends to high variability during the time, thus it might be not efficient to find the best setting of the mobile sampling windows.

\begin{figure}
	\centering
	\includegraphics[width=\linewidth]{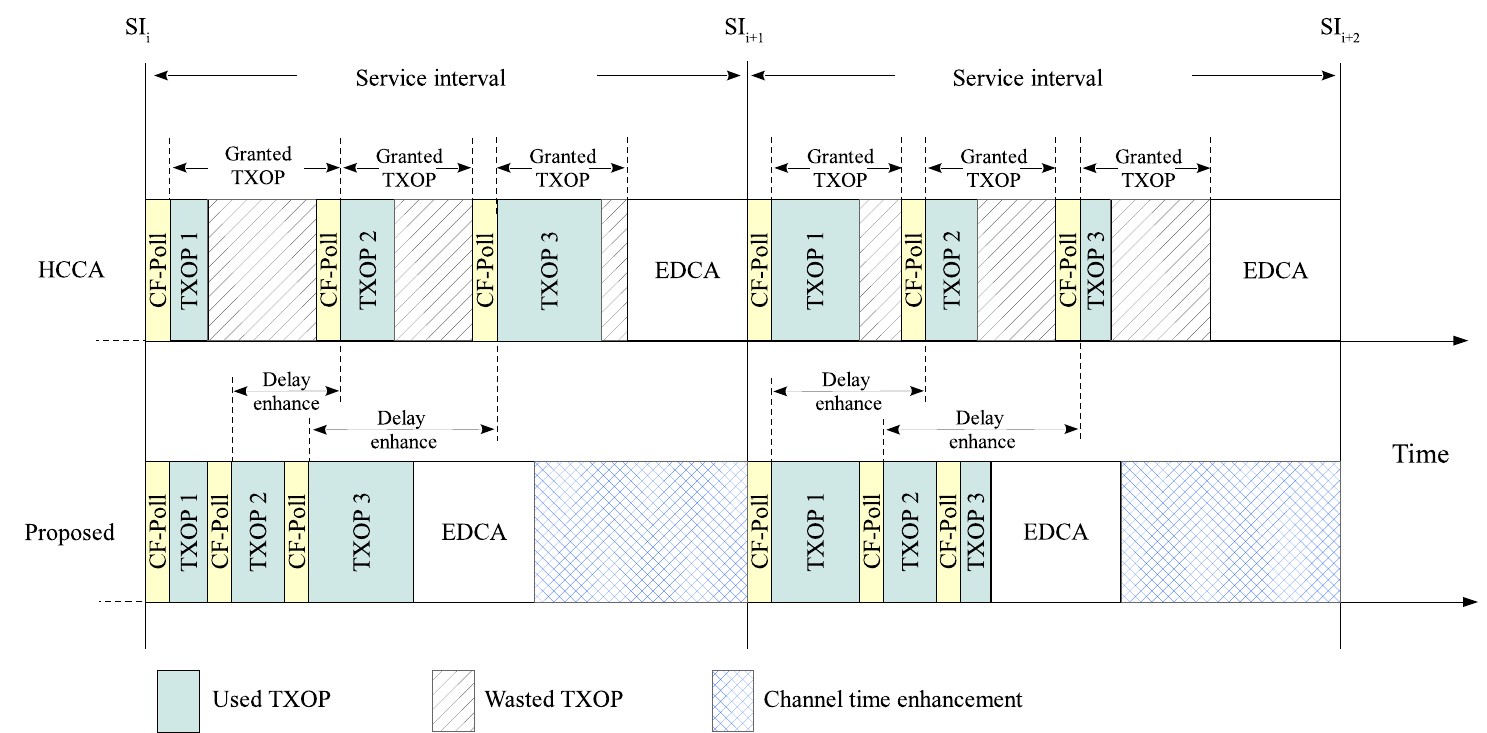}
	\caption{Dynamic TXOP assignment scheduling algorithm.}
	\label{dig:HCCAvsATAV}
\end{figure}

\textbf{The Dynamic TXOP (D-TXOP) scheduling algorithm} \cite{Almaqri2013} analyzes the video of the prerecorded streams before the call setup, which has been previously highlighted in \cite{Haddad2012}. The D-TXOP is suitable  for transmitting type (I) of VBR video source categorized in Subsection~\ref{sec:Adpt2Aplica}, which shows variability in packet size. Indeed, this approach assigns the \gls{TXOP} for a stream based on the real frame size rather than the estimated average of frame size. It uses the unused \gls{QS} field of IEEE802.11e \gls{MAC} header to send the actual size of the upcoming frame to the \gls{HC}. Thus, the wasted TXOPs have been minimized by this approach, which reflects lower delay compared to the previous solutions. Moreover, the EDCA benefits from the surplus TXOP duration from unused TXOP of the preceding STAs as illustrated in Fig.~\ref{dig:HCCAvsATAV}.
\subsection{\gls{HCCA} admission control enhancements}
\label{sec:ACUEnh}
\begin{figure}[!t]
	\centering
	\includegraphics[width=.9\linewidth]{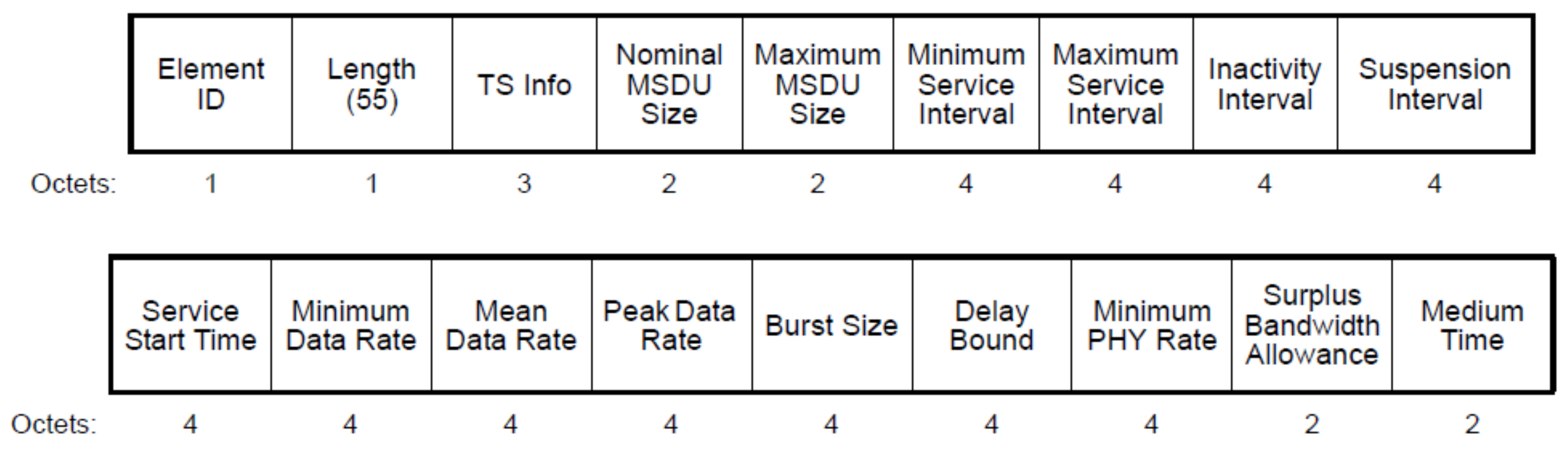}
	\caption{\gls{TSPEC} element format}
	\label{fig:TSPECElements}
\end{figure}

The main purpose of \gls{HCF} admission control is to administer policy or regulate the available bandwidth resources which is used by the \gls{HC}. The admission control is used to limit the amount of traffic admitted under a certain service category in order to guarantee the highest possible QoS level, while maximizing the utilization of the medium resources. Fig.~\ref{fig:TSPECElements} depicts a common frame format for carrying \gls{TSPEC} parameters. Since the admission control relies on a fixed \gls{TSPEC} element, it cannot efficiently cope with the high variability of \gls{VBR} streams. To solve this problem, numerous enhancements and optimizations have been proposed to tackle this deficiency in the legacy \gls{ACU} mechanism.

\textbf{Rate-Variance-envelop-based Admission Control (RVAC)} mechanism uses the \gls{DTB} shaper to guarantee the desired QoS specification \cite{gao2008}. The authors of these two references \cite{ knightly1997, knightly1998} have derived the delay probability based on the aggregate traffic statistics rather than considering each flow individually to accept a new flow for admission \cite{ fan2004, chou2005}. The effective \gls{TXOP} duration of a recently arrived \gls{VBR} stream can be inversely derived based on a given packet loss rate as in Equation~\eqref{eq:ACU}.

Indeed, the RVAC takes the multiplexing gain of \gls{VBR} traffic into account unlike the guarantee-rate-based scheme. More specifically, if the arrival time of data streams extends over a wide range, where the RVAC can fully utilize the multiplexing gain among the \gls{VBR} streams, the performance gain can be noticeable. Additionally, the RVAC considers both uplink and downlink traffic streams. Simulation results have shown that the admission capacity of the RVAC approach is more than the double of its equivalent in the GRAC approach. In addition, the RVAC scheme will not violate the 0.1 second delay requirement as long as the starting time of the streams are spread over a wide time range of not less than 2 seconds. However, the performance of the RVAC in the wireless channel errors environment is not studied.

\textbf{Equal-SP} \cite{zhao2008} has been designed of \gls{HCCA} scheduling, in which the spacing of a particular stream is determined as the period of time between two consecutively scheduled streams. It has been called equal-SP scheduling because a particular stream will always get an equal spacing for its scheduling slices in the schedule. Indeed, the equal-SP scheduling relies on the well-known \gls{RM} algorithm to achieve the QoS requirements. Despite that the equal-SP approach is similar to the SETT-EDD in terms of the general scheduling concept, however, the former assigns equal spacing for each particular stream, which is proven to violate the delay requirement in some cases. 

\begin{figure}[h!]
	\centering
	\includegraphics[width=.8\linewidth]{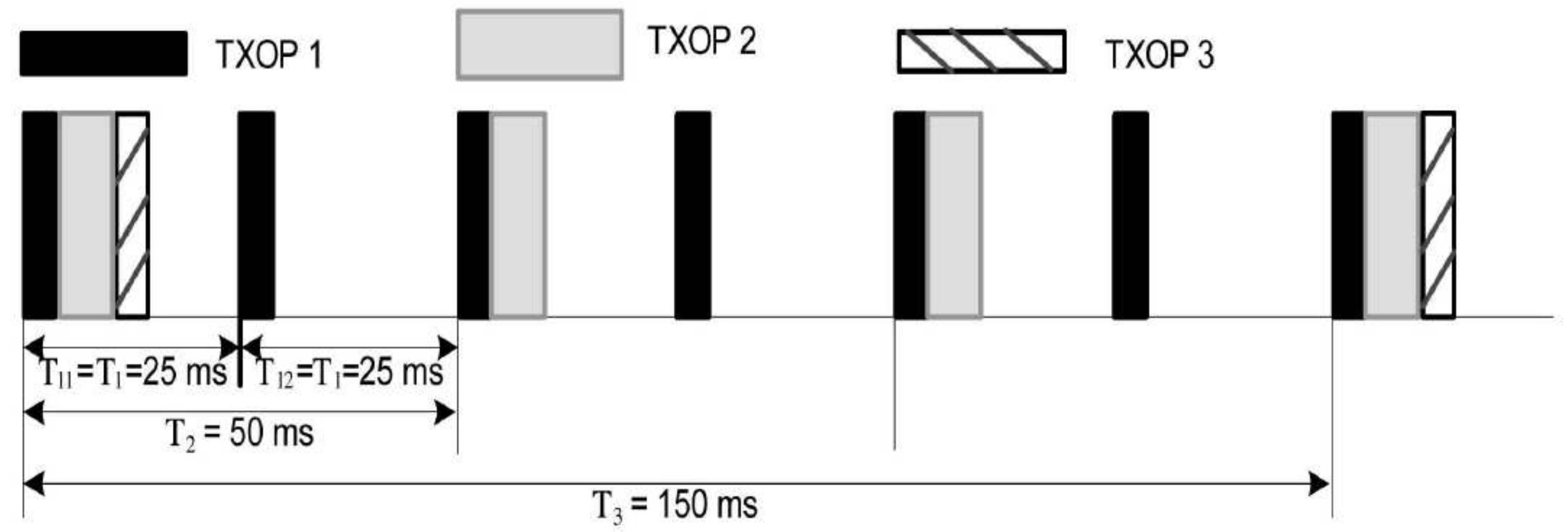}
	\caption{An Example of the Equal-SP scheduling. The QoS is guaranteed by applying admission control}
	\label{fig:EqualSP}
\end{figure}

In the example as shown in Fig.~\ref{fig:EqualSP}, the scheduler assigned 25 ms, 50 ms, and 150 ms time spicing for the flows 1, 2, and 3, respectively, which makes $T_{11}=T_{12}=T_{1}=$25 ms, $T_{2}$=50 ms, and $T_{3}=$150 ms. The equal-SP approach is easy to be implemented and it can guarantee the delay requirements and efficiently utilize the bandwidth while maintaining the compatibility to the standard since it uses the same \gls{TSPEC} parameters. However, the equal-SP approach encounters the same issues faced by the standard; if a newly admitted stream has a smaller delay bound than the current $T_{1}$, the current $T_{1}$ will be set to less than or equal to the new delay bound. Therefore, the \gls{TXOP} durations for the previously admitted flows are required to be recalculated with the $T_{i}$s. Additionally, the scheduler needs to reassign indexes to the admitted flows in order to maintain the condition of \mbox{$T_{1} \leq T_{2} \leq \cdots \leq T_n$}.

\begin{figure}[h!]
	\centering
	\includegraphics[width=.8\linewidth]{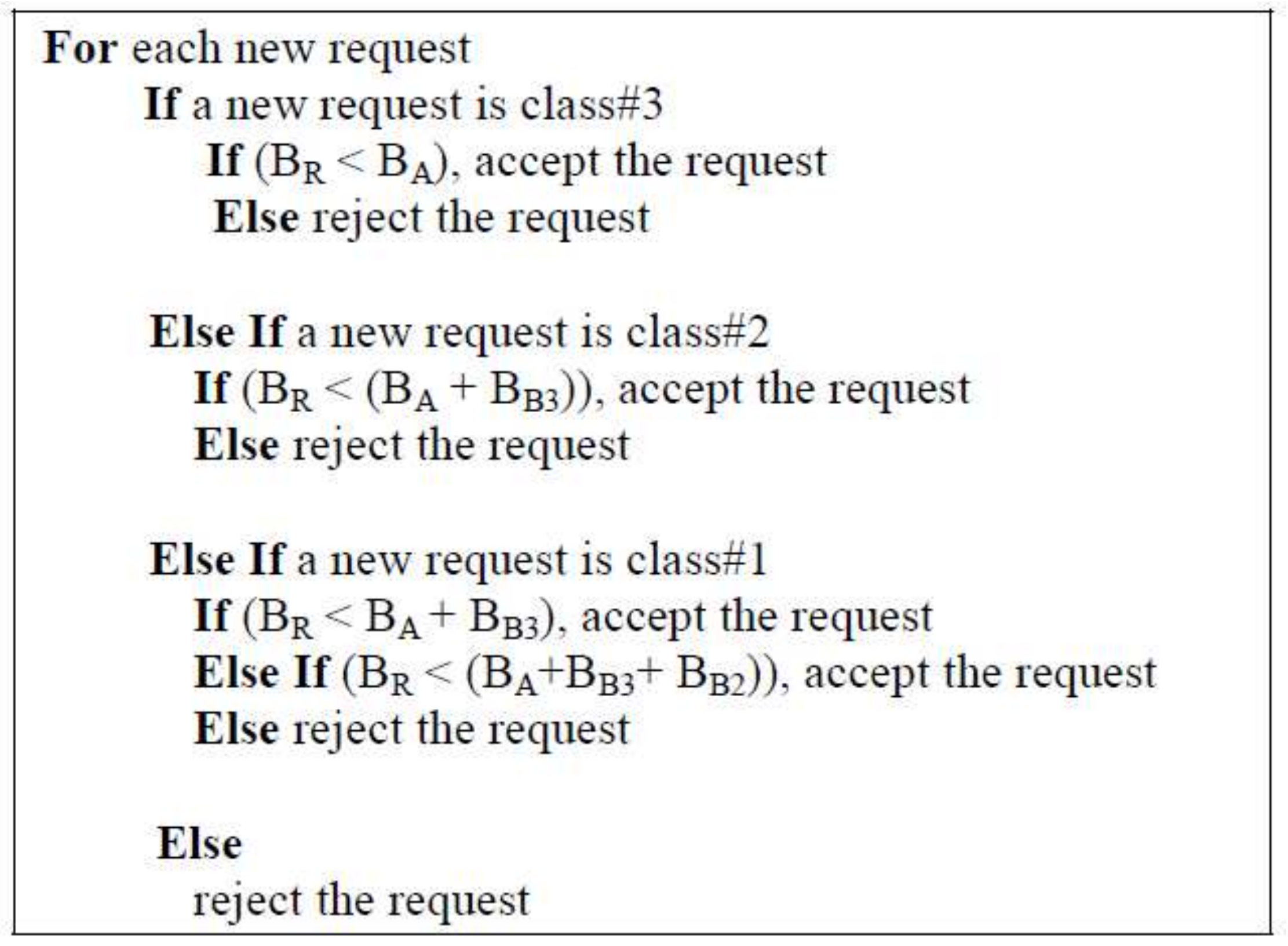}
	\caption{PHCCA admission control mechanism}
	\label{fig:PHCCA}
\end{figure}

\textbf{PHCCA}, as described in Fig.~\ref{fig:PHCCA}, is a priority based QoS and admission control used for queue management mechanism \cite{hantrakoon2010}. In this approach, a  mechanism for borrowing and returning bandwidth among queues has been studied. The higher priority queue, called class, has permission to borrow bandwidth from lower priority queues with the awareness of starvation protection for each priority queue. PHCCA modifies the \gls{HCCA} by classifying the traffic into 3 classes, which has not been divided by the standard. Class 1 is the highest priority class suitable for voice and conference traffic implementation. Class 2 is the second highest priority class suitable for broadcast video traffic. Class 3 is the lowest priority traffic suitable for FTP and HTTP traffic.

Experimental results reveal that the proposed PHCCA can accept more requests for Class 1 (70\% better) compared to the regular \gls{HCCA}. Meanwhile, the second highest priority (Class 2 in this case) is still served quite similar to the regular \gls{HCCA}. Despite that the PHCCA significantly outperforms the regular \gls{HCCA}, it is still not able to guarantee the required QoS for every flow; since it assumes that flows of similar types (e.g. VoIP) will have exact QoS requirements. Besides, the  performance of NPHCCA could merely achieve the performance level of  Best-Effort for VoIP and CBR-video applications. Moreover,  parameters and environment factors for bandwidth borrowing mechanism, such as distance from the access point or mobility, should be investigated.

\textbf{AF-HCCA} \cite{almaqri2017a} enhances the experienced delay of video traffics by utilizing the surplus bandwidth and mitigates the over-polling issue. It computes the TXOP for a traffic stream based on the knowledge about the actual upcoming frame size instead of assigning TXOP according to the mean characteristics of the traffic which is unable to reflect the actual traffic. This scheduler exploits the queue size field of IEEE802.11e MAC header to transfer this information to the HC.
 
In AF-HCCA, the QSTAs will be prevented from receiving unnecessary large TXOP which produces a remarkable increase in the packet delay. Furthermore, the surplus time of the wireless channel conserved by reducing the number of poll frames throughout the feedback is another benefit of this research.The integrated scheme of AF-HCCA shows superior performance compared to IEEE802.11e HCCA, Enhanced EDD \cite{Lee2009} and F-Poll \cite{maqri2015} schedulers in terms of delay and channel utilization without affecting the system throughput. However, preserved TXOP time is not efficiently utilized to enhance the flow capacity.

\textbf{Feedback-based Admission Control Unit (FACU)} \cite{almaqri2017} aims at maximizing the utilization of the surplus bandwidth which has never been tested in previous schemes. The FACU exploits piggybacked information containing the size of the subsequent video frames to increase the number of admitted flows. 

The FACU introduces an enhancement to admission control mechanism of Adaptive-TXOP. Analytical results reveal the efficiency of FACU over the examined schemes. The results show that the conserved channel bandwidth of Adaptive-TXOP can be utilized to increase the number of admitted QoS flows and enhance the overall QoS provisioning in IEEE802.11e WLANs.

\section{\gls{HCCA} scheduling approaches comparison}
\label{sec:comparison}
Table~\ref{tab:comparison} presents a summary and comparison for the \gls{HCCA} enhancements in IEEE802.11e along with their targeted features classified based on the place of the enhancement. The solution column briefly describes the used technique. The complexity of an approach can be high, medium or simple estimated based on Likert-type rating scale. The complexity here represents the volume of the operations of that particular approach. The method that involves more operations is considered high-complex and vice versa. The main targeted traffic of the enhancement is stated. The targeted flow direction, which is considered by the approach, is also presented. 
\begin{table*}[h!] \scriptsize
	\begin{center}
		\caption{Comparison of The main characteristics of the \gls{HCCA} approaches}
		\label{tab:comparison}
		\begin{tabular}{m{2cm} m{2.5cm} m{3.5cm} m{1cm} m{1.5cm} m{1.6cm} m{2.5cm}} \hline
			Strategy  		& Approach 			& Solution  		& Complexity Level$^*$ 		& Targeted & Flow Direction & Main QoS challenge\\\hline
			& CP-Multipoll \cite{Lo2003} 				 & Multipolling scheme 		 				& High   & Voice/video   	& Uplink/downlink & Packet delay\\
			HCCA polling 	& CF-Poll \cite{lee2006} 					 & Piggyback 				 				& Medium & Voice/video   	& Uplink & Flows capacity\\
			mechanisms   	& DEB method for HCCA \cite{Huang2010}  	 & Deterministic polling 	 				& Simple & CBR/VBR video 	& Uplink & Flows capacity\\
			& NPHCCA \cite{Chen2011} 					 & Non polling feedback-based 				& Simple & Voice/video   	& Uplink &  Packet delay\\ 
			& F-Poll \cite{maqri2015} 					 &  Feasible Polling Scheme				& Simple & Video & Uplink & Packet delay and flow capacity\\ \hline
			& SETT-EDD \cite{Grilo2003} 				 & Token bucket and Earliest Due Date based & Medium & Voice and video 	& Uplink/downlink & Packet delay\\
			& AMTXOP \cite{almaqri2016} 				 & Adaptive Multipolling TXOP Scheme & Simple & Video 	& Uplink & Packet delay, flow capacity\\
			TXOP allocation & ARROW \cite{Skyrianoglou2006} 			 & Feedback based 							& Simple & Voice and video 	& Uplink/downlink & Flow capacity\\
			mechanism 		& Enhanced ED \cite{Lee2009} 				 & Estimation and feedback based 			& Medium & Voice and video 	& Uplink & Packet delay\\
			& Dynamic TXOP HCCA \cite{cecchettielAL2012} & Bandwidth reclaiming mechanism 			& High 	 & Voice and video 	& Uplink & Packet delay, flow capacity\\ 
			& D-TXOP \cite{Almaqri2013} 						 & The Dynamic TXOP Scheduling algorithm		& Simple & VBR traffic 		& Uplink & Packet delay\\
			\hline
			& RVAC \cite{gao2008} 						 & Dual token bucket (DTB) shaper 			& Medium & VBR traffic 		& Uplink/downlink & Flows capacity\\
			HCCA admission control	& Equal-SP \cite{zhao2008} 		 	 & Equal spacing scheduling 				& Simple & Voice and video 	& Uplink & Flows capacity\\
			& PHCCA \cite{hantrakoon2010} 				 & Priority based 							& Simple & Voice, video 	& Uplink/downlink & Flows capacity\\ 
			& AF-HCCA \cite{almaqri2017a} 				 & Adaptive Feedback-based HCCA	& Simple & Video 	& Uplink & Packet delay \\			
			& FACU \cite{almaqri2017} 				 & Feedback-based Admission Control Unit	& Simple & Video 	& Uplink & Packet delay, flow capacity \\			
			\hline
		\end{tabular}
		\begin{flushleft}
			\hspace{0.2cm} {\tiny $^*$Note that the complexity level reflects the volume of the operations as explained in Section \ref{sec:comparison} based on Likert-type rating scale}
		\end{flushleft}
	\end{center}
\end{table*}
\section{Open research issues}
\label{sec:ORI}
Although the existing approaches provide several possible solutions to alleviate the deficiency of scheduling for \gls{VBR} multimedia traffic in IEEE802.11e WLANs, many issues have been thoroughly discussed in the literature review section, which are potential research topics. This section highlights the most important issues in order to determine the directions for potential future research. One of the problems with \gls{HCCA} is the coexistence. Several mechanisms claim to be able to coordinate different HCs that operate on the same frequency channel. Since HCCA's QoS guarantee depends on the exclusive usage of the frequency channel, multiple \gls{HCCA} can hardly coexist. On the other hand, additional delay may occur by the polling STAs with scalable video that exhibits constant quality yet introduce high variation in the traffic profile. From the cross-layer perspective and to the best of our knowledge, there is no proactive scheme that provides a good solution to the adequate interaction between the fluctuation of the uplink \gls{VBR} traffic profile at the application layer and the flexible scheduling policy at \gls{MAC} layer which exhibit low-complexity design.
In summary, some issues are needed to be considered to provide optimal enhancement for the transmission of \gls{VBR} traffic in IEEE802.11e WLANs. We believe the following suggestions are desirable for designing a good \gls{HCCA} scheme in IEEE802.11e wireless networks.
\begin{itemize}
	\item Providing efficient estimation of the bandwidth in order to achieve high connection throughput.
	\item Designing a scheme coupled with link adaptation mechanisms in order to provide efficient adaptation to dynamic network behavior.
	\item Exploiting the distributed feature of \gls{EDCA} in designing a hybrid \gls{HCCA} scheme in order to yield high integration and interoperability without jeopardizing the system simplicity.
	\item Enabling the fragmentation and the block acknowledgment introduced in the standard \cite{IEEEStandard2012} with \gls{HCCA} scheme.
	\item Achieving low algorithm complexity.
\end{itemize}
\section{Research trend on QoS support in IEEE802.11e}
\begin{figure}
	\centering
	\includegraphics[width=0.8\linewidth]{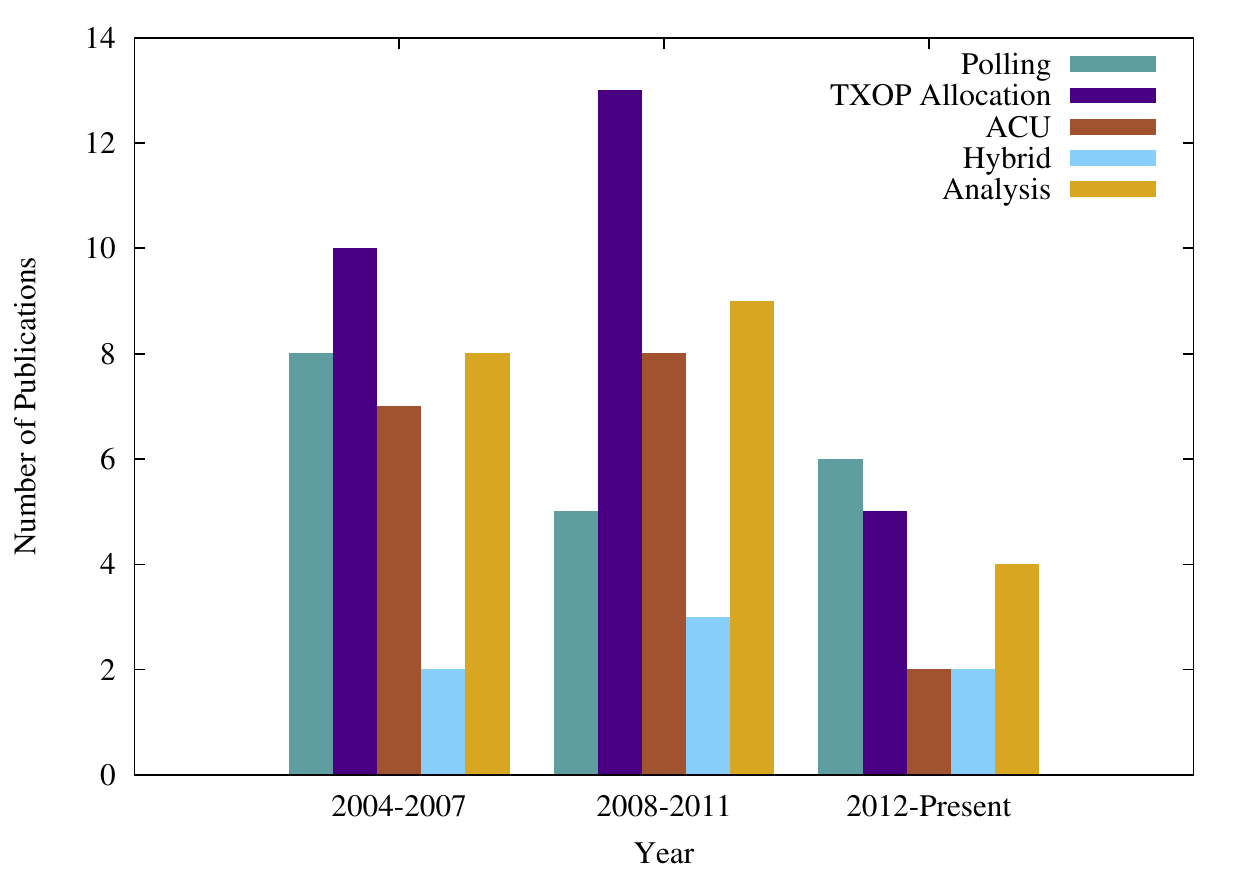}
	\caption{Number of publication of the investigated research areas}
	\label{fig:ResearchTrends}
\end{figure}

Many researches have been conducted in the Literature since the first advent of the HCCA protocol draft in IEEE802.11e standard \cite{IEEE80211eDraft}. These researches can be classified into five research areas as in Table~\ref{tab:ResearchTrend} aims at demonstrating the trend of the research since the first presence of the HCF functions till 2009 and from 2010 to present. The collection includes over 89 journal and conference papers. These scientific documents have been collected using IEEE Explorer Digital Library, Springer Link, ScienceDirect and Google Scholar. One can notice that the polling  and TXOP allocation mechanisms have  greatly received the researchers' attention since the evolution of the HCCA till now, while admission control mechanisms have less interest. It is worth noting that the design of the hybrid EDCA-HCCA scheme has scarcely studied. The HCCA performance and mathematical analysis have been fairly covered. Yet, only few efforts have focused on designing a comprehensive analytical model for HCCA protocol. The aggregated number of papers published in three periods, namely 2004 to 2007; 2008 to 2011 and 2012 to present are depicted in Fig.~\ref{fig:ResearchTrends}. The figure shows that the polling and TXOP mechanism have received a great amount of attention compared to ACU and hybrid scheme. On the other hand, recently there has been a few analysis of HCCA protocol, in contrast to the period from 2004 up to 2011.

\begin{center}
	\begin{table*}[!t]
		\centering
		\caption {Researches in QoS provisioning of Multimedia traffic in IEEE802.11e }
		\centering
		\begin{tabular}{l|l|l}
			\hline
			Area & Published from 2004 to 2009 & Published since 2010\\ \hline
			
			HCCA Polling & 
			\cite{kim2004,Son2004,Xiyan2004,Chen2004,ByungSeoKim2005,lee2006,Ramos2006,Park2007,Chen2008} & \cite{Huang2010,He2011,Chen2011,Chou2011,Viegas2012,zhang2013,Li2013,ZiTsan2014,maqri2015,Zhang2015} \\ \hline
			TXOP Allocation & \cite{Ansel2004,Inan2006,Kim2006,Yamane2006,Skyrianoglou2006,Ansel2006,Choi2007,Qinglin2007,Rashid2007,Kyung2007}, & \cite{Byung2010,noh2010,arora2010,ruscelli2011,luo2011cross,Jansang2011,Yong2011,noh2011,cecchettielAL2012,Cecchetti2012} \\ 
			& \cite{Floros2008,rashid2008,Hsieh2008,Huang2008,Chu2008} & \cite{ju2013,lee2013,almaqri2016} \cite{almaqri2017a}\\ \hline
			
			Admission Control & 
			\cite{fan2004,Deyun2005,Deyun2005a,Kim2006,vander2006,Jang2006,Cecchetti2007,gao2008,Qinglin2008,Zeng2008} & \cite{lee2010,Didi2010,hantrakoon2010,palirts2010,Chie2011,lee2013,Huang2015}, \cite{almaqri2017} \\ \hline
			
			Hybrid EDCA-HCCA & 
			\cite{xiao2004,Fallah2007,Fallah2008,Zhu2008109} & 
			\cite{Siddique2010,ruscelli2012,ng2013} \\ \hline
			
			HCCA Analysis & \cite{Boggia2005325,Qiang2005per,Tresk2006,Harsha2006,Karanam2006,rashid2006,Siris2006,binMuhamad2007,rashid2008,Cecchetti2008}, & \cite{Perez2010,Minseok2010,Ghazizadeh2010,lyakhov2011,Pastrav2012,Lagkas2013,leonovich2013,Jansang2013}\\ 
			& \cite{Ghazizadeh2008,Lin2008,Jansang2009}& \\ \hline
		\end{tabular}
		\label{tab:ResearchTrend}
	\end{table*}
\end{center}

\section{Future directions}
\label{sec:futureWork}
Although all proposed schemes in their current states improve the transmission of prerecorded video, there still some issues need to be addressed and investigated. Below, we highlight some future works that need to be carried out for further enhancement to:
\begin{itemize}
	\item Enhance the HCCA to cope with more complicated wireless scenarios, where the hidden node problem exists and QSTAs communicate using RTS/CTS mechanism with \gls{MAC} level fragmentation and multi-rate support enabled.
	\item Study the scalable \gls{HCCA} \gls{MAC} for video over wireless mesh networks that are also scalable to a wider range of \gls{MAC} settings to support more robust time-bounded media applications.
	\item Design a new admission control algorithm to utilize the excess bandwidth and to manage the available resources among the \gls{HCCA}, \gls{HCF} and \gls{EDCA} in order to maximize the number of served streams or applications in the network.
	\item Examine the performance of HCCA with the presence of collision occurred in the overlapping area when polling stations among two neighboring \glspl{BSS} simultaneously.
\end{itemize}

\section{Conclusion}
\label{sec:conclusion}
IEEE802.11e is aimed at providing stringent QoS support to multimedia applications such as video streaming. The controlled based function of IEEE802.11e standard, \gls{HCCA} scheduler, consider a fixed \gls{TSPEC} for scheduling the traffic while in fact the \gls{VBR} traffic tends to change their characteristics such as data rate and packet size over the time. The inability of the IEEE802.11e \gls{MAC} protocol to accommodate to the high fluctuation of VBR video profile motivates many researches to be conducted. Several enhancements have been made to alleviate these shortcomings. These enhancements tend to address particular issues or applications by improving, in most cases, one of the \gls{HCF} functions. In general, designing a robust \gls{HCF} solution that provides an integrated solution for all traffic classes is still a challenging task for future research.


\begin{IEEEbiography}[{\includegraphics[width=1in,height=1.25in,clip,keepaspectratio]{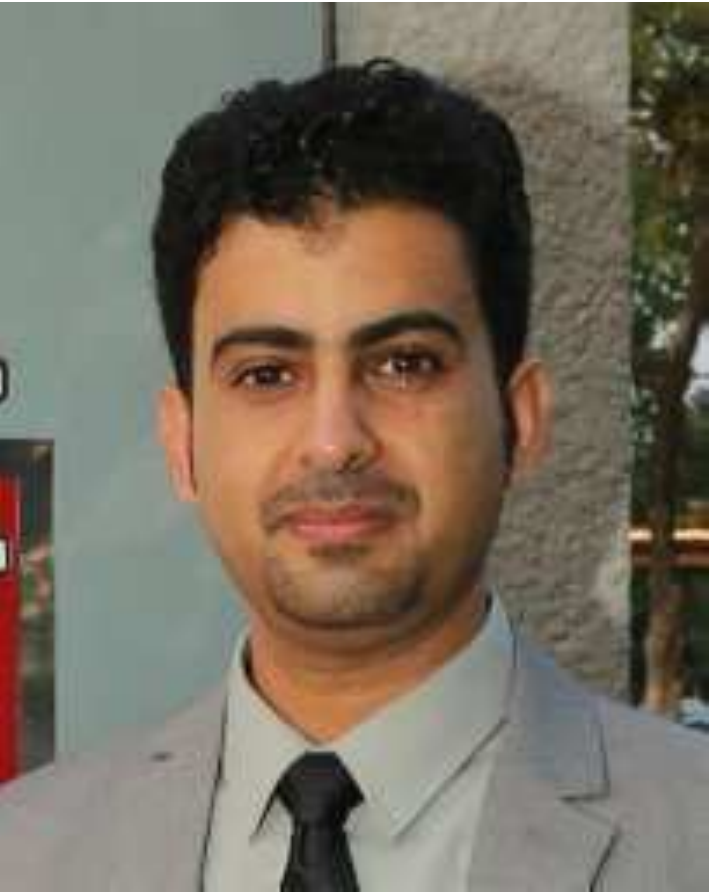}}]{Mohammed A. Al-Maqri}
	received his BSc degree in Computer Science from Almustanseriah University - Iraq, in 2004. Then, he received his MSc and Ph.D degrees in communication technology and networks from Universiti Putra Malaysia in 2009 and 2016, respectively. Now, he is a Lecturer and Head of department of Information Technology in the Faculty of Computer Science and Information Technology, Azal University for Human Development, Sana'a, Yemen. He has published a number of articles in high-impact factor scientific journals. His research interests are in the field of high-speed TCP protocols, high-speed network, QoS, scheduling algorithms, admission control and wireless networks.
\end{IEEEbiography}

\begin{IEEEbiography}[{\includegraphics[width=1in,height=1.25in,clip,keepaspectratio]{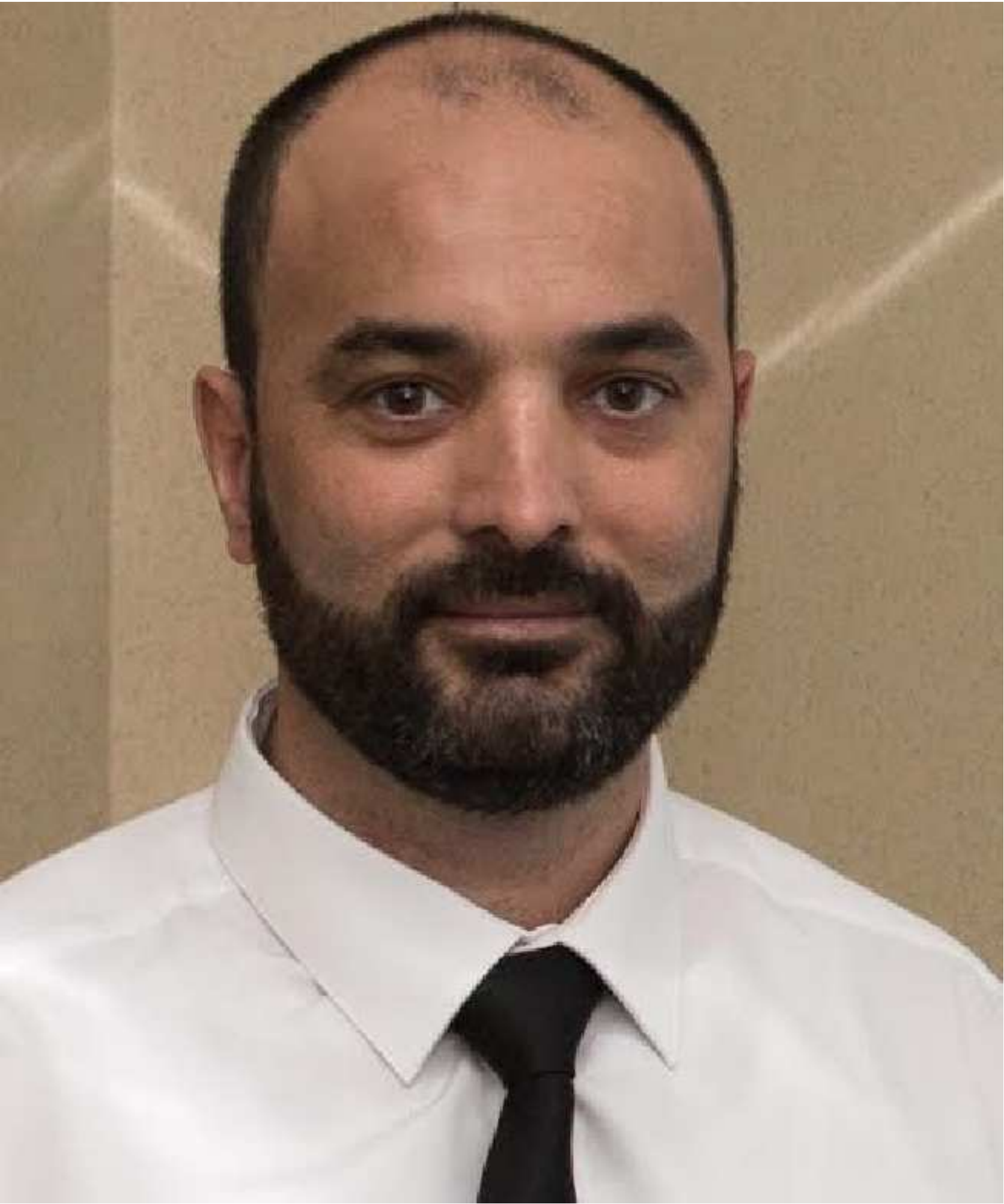}}]{Mohamed A. Alrshah}
	(M'13--SM'17) received his BSc degree in Computer Science from Naser University - Libya, in June 2000. Then, he received his MSc and Ph.D degrees in communication technology and networks from Universiti Putra Malaysia (UPM) in May 2009 and Feb 2017, respectively. Now, he is a Senior Lecturer in the Department of Communication Technology and Networks, Faculty of Computer Science and Information Technology, Universiti Putra Malaysia (UPM). He has published a number of articles in high-impact factor scientific journals. His research interests are in the field of high-speed TCP protocols, high-speed wired and wireless network, WSN, MANET, VANET, parallel and distributed algorithms, IoT and cloud computing.
\end{IEEEbiography}
\newpage

\begin{IEEEbiography}[{\includegraphics[width=1in,height=1.25in,clip,keepaspectratio]{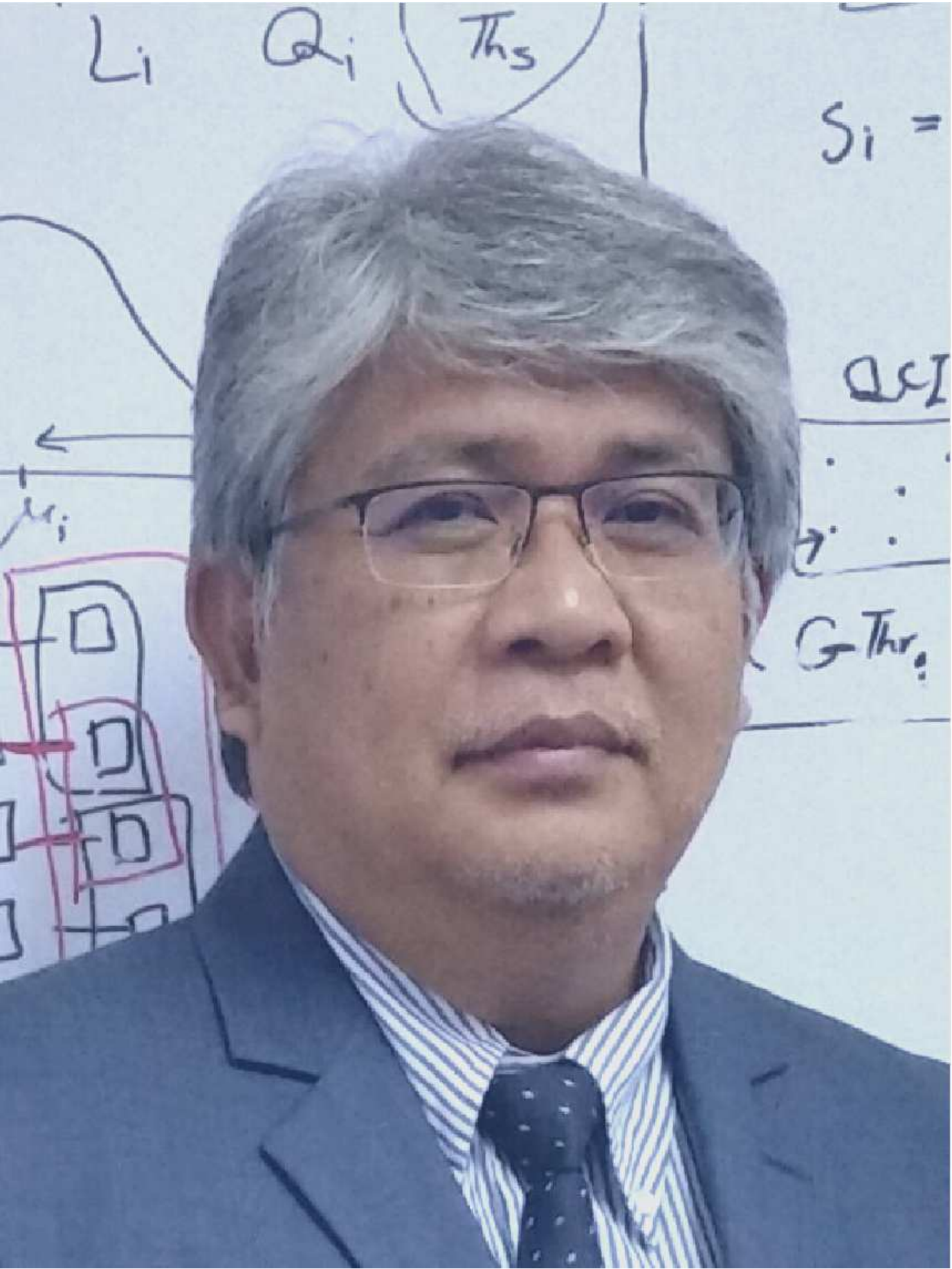}}]{Mohamed Othman}
	(M'06--SM'18) received his Ph.D from the Universiti Kebangsaan Malaysia (UKM) with distinction (Best Ph.D Thesis in 2000 awarded by Sime Darby Malaysia and Malaysian Mathematical Science Society). Now, he is a Professor in the Department of Communication Technology and Networks, Faculty of Computer Science and Information Technology, Universiti Putra Malaysia (UPM). He is also an associate researcher at the Lab of Computational Science and Mathematical Physics, Institute of Mathematical Research (INSPEM), UPM. He published more than 160 International journals and 230 proceeding papers. His main research interests are in the fields of high-speed network, parallel and distributed algorithms, software defined networking, network design and management, wireless network (MPDU- and MSDU-Frame aggregation, MAC layer, resource management, and traffic monitoring) and scientific telegraph equation and modelling.
\end{IEEEbiography}
\EOD

\vfill
\end{document}